\pdfoutput=1 
\documentclass[a4paper, 11pt]{article}
\PassOptionsToPackage{table}{xcolor}
\usepackage{cancel}

\usepackage{jheppub} 

\usepackage[T1]{fontenc} 
\usepackage{tangocolors}
\usepackage{amsmath}
\usepackage{tikz}
\usetikzlibrary{arrows, decorations,backgrounds, patterns}
\usetikzlibrary{decorations.pathreplacing ,decorations.markings}
\usepackage{amssymb}\usepackage{amsfonts}
\usetikzlibrary{decorations.pathmorphing,backgrounds,shapes,arrows,shadows}
\usetikzlibrary{patterns.meta}
\usetikzlibrary{patterns,decorations.pathmorphing}
\tikzset{
    snake it/.style={decorate, decoration=snake}
}
\usepackage{pgfplots}
\pgfplotsset{compat=1.11}
\usepgfplotslibrary{fillbetween}
\usetikzlibrary{intersections}
\pgfdeclarelayer{bg}
\pgfsetlayers{bg,main}
\tikzset{zigzag/.style={decorate,decoration=zigzag}}
\tikzset{snake it/.style={decorate, decoration=snake}}
\makeatletter
\def\@hex@@Hex#1%
 {\if a#1A\else \if b#1B\else \if c#1C\else \if d#1D\else
  \if e#1E\else \if f#1F\else #1\fi\fi\fi\fi\fi\fi \@hex@Hex}
\makeatother

\usetikzlibrary{arrows.meta}
\tikzset{ mid arrow/.style={ postaction={decorate}, decoration={ markings, mark=at position 0.6 with {\arrow{#1}} } } }
\tikzset{>={Latex[width=2mm,length=2mm]}}

\usepackage[all]{xy}
\DeclareFontEncoding{LS2}{}{\noaccents@}
\DeclareFontSubstitution{LS2}{stix}{m}{n}

\DeclareSymbolFont{integrals}{LS2}{stixcal}{m}{n}

\DeclareMathSymbol{\ointupbig}{\mathop}{integrals}{"E8}
\DeclareMathSymbol{\ointupsmall}{\mathop}{integrals}{"B2}

\usepackage[percent]{overpic}
\usepackage{slashed}
\usepackage{wrapfig}
\usepackage{tabu}
\usepackage{diagbox}
\usepackage{mathrsfs,amsmath,amssymb,amsthm,amsfonts,tikz,graphicx,accents,hyperref, color}
\usepackage{dsfont,epiolmec, latexsym, stmaryrd, comment}
\usepackage{slashed,ccaption}
\usepackage{mathrsfs, calligra}
\usepackage{leftidx}
\usepackage{import}
\usepackage{multirow}
\usepackage{amsfonts}
\usepackage{pifont}
\usepackage{tabularx}
\usepackage{cancel}
\usepackage[utf8]{inputenc}
\usetikzlibrary{intersections,calc}
\usepackage{ifthen}
\usepackage{amsmath}
\usepackage{cancel}
\usepackage{caption} 
\usepackage{subcaption}

\usetikzlibrary{patterns.meta}
\usepackage{array}
\hypersetup{ linktoc=all,
    colorlinks, linkcolor={palatinateblue},
    citecolor={brightpink}, urlcolor={amaranth}
}

\graphicspath{{Images/}}

\renewcommand{\d}[1]{\ensuremath{\operatorname{d}\!{#1}}}
\def\bohm{\boldsymbol{\mho}}

\def\sideremark#1{\ifvmode\leavevmode\fi\vadjust{\vbox to0pt{\vss
 \hbox to 0pt{\hskip\hsize\hskip1em
 \vbox{\hsize2cm\tiny\raggedright\pretolerance10000
 \noindent #1\hfill}\hss}\vbox to8pt{\vfil}\vss}}}%
\DeclareSymbolFont{extraup}{U}{zavm}{m}{n}
\DeclareMathSymbol{\varheart}{\mathalpha}{extraup}{86}
\DeclareMathSymbol{\vardiamond}{\mathalpha}{extraup}{87}
\makeatletter
\renewcommand*{\@fnsymbol}[1]{\ensuremath{\ifcase#1\or \clubsuit \or \vardiamond \or \varheart\or
    \spadesuit\or \mathparagraph\or \|\or **\or \dagger\dagger
    \or \ddagger\ddagger \else\@ctrerr\fi}}
\makeatother

\definecolor{rosy}{RGB}{230,235,252}
\definecolor{myframetitle}{RGB}{90,89,170}
\definecolor{myblocktitle}{RGB}{140,185,249}
\definecolor{mytitle}{RGB}{10,80,26}

\definecolor{darkgreen}{RGB}{27,130,45}
\definecolor{darkblue}{rgb}{0,0,0.3}
\definecolor{darkred}{rgb}{0.7,0,0}

\definecolor{light gray}{RGB}{220,220,220}
\definecolor{dark purple}{RGB}{108,0,217}
\definecolor{pink}{RGB}{190,20,100}
\definecolor{orang}{RGB}{193,63,0}
\definecolor{green}{RGB}{11,98,17}
\definecolor{darkpink}{RGB}{153,0,76}
\definecolor{bluegreen}{RGB}{0,102,102}
\definecolor{greenlagan}{RGB}{0,102,0}
\definecolor{redgreen}{RGB}{102,102,0}
\definecolor{Redgreen}{RGB}{153,76,0}
\definecolor{vividviolet}{rgb}{0.62, 0.0, 1.0}
\definecolor{amaranth}{rgb}{0.9, 0.17, 0.31}
\definecolor{palatinateblue}{rgb}{0.15, 0.23, 0.89}
\definecolor{brightpink}{rgb}{1.0, 0.0, 0.5}
\definecolor{cornflowerblue}{rgb}{0.39, 0.58, 0.93}
\definecolor{deepcarminepink}{rgb}{0.94, 0.19, 0.22}
\definecolor{radicalred}{rgb}{1.0, 0.21, 0.37}

\newcommand\hnote[1]{\textcolor{blue}{\bf [HA:\,#1]}}
\newcommand\snote[1]{\textcolor{darkpink}{\bf [Sh:\,#1]}}
\newcommand\vnote[1]{\textcolor{cyan}{\bf [V:\,#1]}}

\newcommand{\bTh}{\boldsymbol{\Theta }}

\newcommand{\bO}{\boldsymbol{\Omega}}

\def\bGIJ{\boldsymbol{{\cal G}}_{_{\mathbb{IJ}}}}
\def\bGJI{\boldsymbol{{\cal G}}_{_{\mathbb{JI}}}}
\def\bG{\boldsymbol{{\cal G}}}
\def\bohm{\boldsymbol{\mho}}

\DeclareFontFamily{OT1}{rsfs}{}

\DeclareFontShape{OT1}{rsfs}{m}{n}{ <-7> rsfs5 <7-10> rsfs7 <10->rsfs10}{} 

\DeclareMathAlphabet{\mycal}{OT1}{rsfs}{m}{n}

\newcommand{\be}{\begin{equation}}
\newcommand{\ee}{\end{equation}}
\newcommand{\bea}{\begin{eqnarray}}
\newcommand{\eea}{\end{eqnarray}}
\usepackage{enumitem}
\makeatletter \@addtoreset{equation}{section}

\newcommand\tcr{\textcolor{red}}

\newcommand\tcbg{\textcolor{bluegreen}}

\begin{document}


\newcommand{\mytitle}{\begin{center}{\LARGE{\textbf{  Carrollian Structure of the Null Boundary Solution Space }}} \end{center}}

\title{{\mytitle}}

\author[a,b]{H.~Adami}
\author[c]{ A.~Parvizi}
\author[c]{, M.M.~Sheikh-Jabbari}
\author[c,d]{, V.~Taghiloo}
\author[b]{, H.~Yavartanoo}
\affiliation{$^a$ Yau Mathematical Sciences Center, Tsinghua University, Beijing 100084, China}
\affiliation{$^b$ Beijing Institute of Mathematical Sciences and Applications (BIMSA), \\ Huairou District, Beijing 101408, P. R. China}
\affiliation{$^c$ School of Physics, Institute for Research in Fundamental
Sciences (IPM),\\ P.O.Box 19395-5531, Tehran, Iran}
\affiliation{$^d$ Department of Physics, Institute for Advanced Studies in Basic Sciences (IASBS),\\ 
P.O. Box 45137-66731, Zanjan, Iran}

\emailAdd{
hamed.adami@bimsa.cn, 
a.parvizi@ipm.ir,
jabbari@theory.ipm.ac.ir, v.taghiloo@iasbs.ac.ir
}

\abstract{We study pure $D$ dimensional Einstein gravity in spacetimes with a generic null boundary. We focus on the symplectic form of the solution phase space which comprises a $2D$ dimensional boundary part and a $2(D(D-3)/2+1)$ dimensional bulk part. The symplectic form is the sum of the bulk and boundary parts, obtained through integration over a codimension 1 surface (null boundary) and a codimension 2 spatial section of it, respectively. Notably, while the total symplectic form is a closed 2-form over the solution phase space, neither the boundary nor the bulk symplectic forms are closed due to the symplectic flux of the bulk modes passing through the boundary. Furthermore, we demonstrate that the $D(D-3)/2+1$ dimensional Lagrangian submanifold of the bulk part of the solution phase space has a Carrollian structure, with the metric on the $D(D-3)/2$ dimensional part being the Wheeler-DeWitt metric, and the Carrollian kernel vector corresponding to the outgoing Robinson-Trautman gravitational wave solution.}
\maketitle
\section{Introduction}\label{sec:Intro}

Question of field theories on spacetimes with boundaries appears in various areas of physics. In particular, the boundary can be a codimension 1 null or timelike surface. It is well known that consistency of the theory besides the usual bulk modes requires introduction of boundary degrees of freedom (dof). These dof reside on the boundary while interacting with the bulk modes and among themselves. {Furthermore,}  {in} theories with local (gauge) symmetries, it is known that the boundary dof {can} be labeled and studied {by}  extending the notion of gauge invariance in {the} presence of the boundary. Explicitly, residual gauge symmetries and the {corresponding} surface charges \cite{Brown:1986nw,Barnich:2001jy,Harlow:2019yfa,Grumiller:2016pqb,Henneaux:2018gfi,Barnich:2009se,Barnich:2011mi,Campiglia:2020qvc,Compere:2019bua,Fiorucci:2020xto,Ashtekar:1996cd,Fuentealba:2023hzq,Klinger:2023qna} provide a systematic {framework} to formulate boundary dof. 

In this work, {our primary emphasis lies in}  $D$ dimensional pure Einstein gravity {in} presence of a null boundary. This problem is motivated by questions {within the realm of} black hole physics, where the horizon plays the role of the null boundary from the viewpoint of non-free-fall, fiducial, observers situated ``outside'' the horizon. The same problem arises when exploring gravity in an asymptotic flat spacetime, where the null boundary is asymptotic future (or past) null infinity. While our main focus will be on the former, the latter has been subject of intense study in the recent years, particularly in connection with memory effects and gravitational wave observations \cite{Strominger:2014pwa,Strominger:2017zoo,Pasterski:2015tva,Blanchet:2023pce,Seraj:2022qyt}. 

In the analysis of asymptotic or boundary symmetries, the conventional approach typically commences with specifying/prescribing the falloff behavior of the fields near the boundary, as demonstrated in, for example, \cite{Penrose:1962ij, Newman:1962cia, Brown:1986nw, Barnich:2009se, Barnich:2010eb, Barnich:2011mi, Strominger:2017zoo, Pasterski:2021rjz, Fuentealba:2021yvo, Prema:2021sjp, Compere:2019bua, Compere:2020lrt, Geiller:2022vto, Godazgar:2020gqd,Compere:2013bya,Perez:2016vqo,Barnich:2016rwk,Oliveri:2019gvm,DePaoli:2018erh,Chrusciel:2023umn,Odak:2021axr,Skenderis:2000in,McNees:2023tus,Shi:2020csw} and \cite{Donnay:2015abr, Adami:2020amw, Grumiller:2019fmp}. However,  in general one may formulate the problem in a different way: One can start with constructing the set of all solutions to the theory which accommodate presence of the boundary. For the case of a gravity on a spacetime with a null boundary (horizon), this program was outlined in \cite{Grumiller:2020vvv} and worked through in \cite{Adami:2020amw, Adami:2020ugu, Adami:2021sko, Adami:2021nnf, Adami:2021kvx, Sheikh-Jabbari:2022mqi}, see also \cite{Hopfmuller:2016scf, Chandrasekaran:2018aop, Chandrasekaran:2021hxc, Chandrasekaran:2023vzb} for related work.

In particular, a complete null boundary solution space for $D$ dimensional Einstein gravity has been constructed in \cite{Adami:2021nnf}. We {place} the null boundary at $r=0$ and construct the solution space by solving {the} Einstein equations {perturbatively}  around $r=0$. This solution space, which we will briefly review in section \ref{sec2:soln-space-review}, 
is {characterized} by $D+ D(D-3)/2$ functions over the $D-1$ dimensional null boundary ${\cal N}$. 

\begin{figure}[t]
\def \L {3.0}
    \centering
\begin{tikzpicture}
  \draw[thick,red] (-0.6*\L,-0.6*\L) coordinate (b) -- (0.9*\L,0.9*\L) coordinate (t);
\draw[thick,green,->] (-0.56*\L,-0.6*\L) coordinate (b) -- (0.94*\L,0.9*\L) coordinate (t);
  \draw[blue,->] (1.7*\L,-0.2*\L) -- (0.6*\L,0.9*\L);           
  \draw[blue,->] (1.6*\L,-0.3*\L) -- (0.4*\L,0.9*\L);           
  \draw[blue,->] (1.5*\L,-0.4*\L) -- (0.2*\L,0.9*\L);
  \draw[blue,->] (1.3*\L,-0.6*\L) -- (-0.2*\L,0.9*\L);
  \draw[blue,->] (1.2*\L,-0.7*\L) -- (-0.4*\L,0.9*\L);
  \draw[black,thick,->] (-0.1*\L,-0.1*\L)--(-0.09*\L,-0.09*\L); \draw[black,thick,->] (-0.08*\L,-0.08*\L)--(-0.07*\L,-0.07*\L);
  \draw[red] (-0.08*\L,0.04*\L) node[left, rotate=45] (scrip) {{$v$}};
  \draw[blue] (1.15*\L,-0.25*\L) node[left, rotate=-45] (scrip) {\small{ infalling null rays}};
  \draw[red] (-0.4*\L,-0.2*\L) node[left, rotate=45] (scrip) {\small{$r=0$}}; 
  \draw[red] (-0.3*\L,-0.45*\L) node[left, rotate=45] (scrip) {\small{${\cal N}$}};
  \draw[green] (2.04*\L,0.8*\L) node[left ] (scrip) {\small{$\text{outgoing RT mode}$}};
  \draw[brown] (0.4*\L,-0.1*\L) node[left] (scrip) {\small{$r>0$}};
  \draw[brown] (0*\L,0.2*\L) node[left] (scrip) {\small{$r<0$}};
\end{tikzpicture}
\caption{Depiction of a null boundary at $r=0$, {spanned by}  $v$ and $x^A$ (perpendicular to the plane) coordinates. Our null boundary solution space {encompasses} all solutions to pure $D$ dimensional Einstein gravity in {the region} $r\geq 0$. {In our notation, the} infalling gravitational waves correspond to ${\cal N}_{AB}$ modes and the RT mode  {denoting the} Robinson-Trautman gravitational wave solution \cite{Robinson:1960zzb, Robinson:1962zz}. {In addition to }  the ``bulk modes'' (RT and infalling modes) we {also} have boundary modes  residing on ${\cal N}$. These modes are specified by $D$ functions at an arbitrary given $v=v_b$.
}\label{Fig:Null-surface-flux-Horizon}
\end{figure}
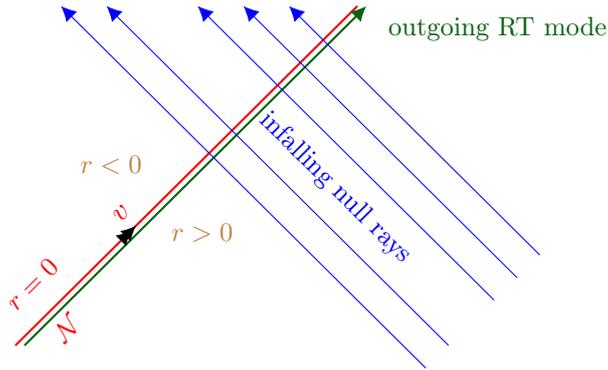

Before {we delve} further, let us introduce a  terminology {that proves particularly useful}  in the {context of} light-front field theory formulation and when {dealing with} a null boundary. As depicted in Fig. \ref{Fig:Null-surface-flux-Horizon},  the null boundary ${\cal N}$ {is} spanned by $v$ (parametrizing the null direction) and $D-2$ dimensional spatial coordinates $x^A$. In light-front field theories, ${\cal N}$ plays the role of partial Cauchy surface {where}  the dynamics means evolution in the light-cone (null) time direction, $v$, and {the conjugate} momentum is related to the derivative of the fields w.r.t $v$. If $\Phi(r;v,x^A)$ denotes a generic field, the solution space is spanned by $\Phi(r=0;v,x^A)$ and the momentum conjugate to $\Phi$, which is proportional to $\partial_v \Phi$, does not carry an independent information. {Consequently,} unlike the usual spatial (partial) Cauchy surfaces, in the Hamiltonian formulation of light-front field theories, half of Hamilton's equations, which define conjugate momenta, simply {represent relationships}  among Cauchy data. {Noting}  this peculiar feature, it {proves}  useful to distinguish between the ``solution space'' and ``solution phase space'' in the light-front field theories  {and when} formulating the theory with a null boundary. If the latter space is $2N$ dimensional, the former is $N$ dimensional and is obtained by imposing half of Hamilton's equations on the fields.  In {essence}, the equations defining conjugate momenta may be {regarded} as (second class) constraints on the system, and the solution space {serves as } the ``reduced phase space'' of the theory. {For further details, refer to Appendix \ref{appen-A}}. A {notable} feature of the light-front solution space is that {in addition to its} symplectic structure {it possesses}  a metric structure. In simple field theories,  {such as} the one discussed in appendix \ref{appen-A}, the metric on the field space also appears in the symplectic form. {Consequently}, the symplectic form on the solution space contains information about the metric on the solution space.

In our analysis, {the primary} focus {is on} the symplectic form of the theory. {We begin with the} symplectic form on the solution phase space, and {by} imposing the boundary equations of motion and definition of conjugate light-cone momenta, we {derive} the on-shell symplectic form, which {pertains to} the symplectic form over the solution space. This {on-shell} symplectic form {consists of} two parts, a boundary part given by integrals over constant $v$ slice at ${\cal N}$ and the bulk part {determined} by codimension 1 integrals over ${\cal N}$. The on-shell boundary part {comprises} $D$ modes and in the absence of bulk modes (when there {are} no gravitational waves passing through the boundary), {it} is closed and invertible. In {such} cases, one can invert the symplectic form, {calculate}  Poisson brackets over the solution space (which are Dirac brackets over the solution phase space), and obtain {the} algebra of surface charges. In a {suitable} slicing of the solution space, this is a direct sum of Heisenberg and Diff($D-2$) algebras \cite{Adami:2021nnf, Adami:2021kvx, Sheikh-Jabbari:2022mqi}. When we turn on bulk modes, neither the boundary nor bulk parts of the on-shell symplectic forms are individually closed or invertible. This feature is anticipated due to the flux of bulk modes {passing} through the boundary. 

The symplectic form on a sector of solution phase space ({referred to as the} off-shell symplectic form) and the {corresponding} Poisson/Dirac brackets, {were} recently studied in \cite{Ciambelli:2023mir}. {In this work}, we {narrow our} focus on the bulk part of the on-shell symplectic form. {To streamline our} analysis,  we consider the co-rotating case, reducing the number of boundary modes to two.  Interestingly, one of the modes, {specifically} the volume form over the codimension 2 spacelike surface spanned by $x^A$, appears {both} as a boundary mode and a bulk mode. {Consequently}, we have $D(D-3)/2+1$ bulk modes, {which is} one more than the {anticipated}  number of gravitational wave polarizations. {Our analysis reveals that} this space, at any given point on ${\cal N}$, possesses a $(D-1)(D-2)/2$ dimensional Carrollian structure. The metric on this Carrollian geometry is the Wheeler-DeWitt metric, {with} the kernel vector {aligning} {with} the mode {shared by both the}  boundary and bulk parts.

\paragraph{Organization of the paper.} In Section \ref{sec2:soln-space-review}, we provide an overview of the construction of the null boundary solution space, {which also establish}  the conventions and notations {used in this work}. In Section \ref{sec:symplectic-form}, we {introduce the}  off-shell symplectic form and the basic Poisson brackets on the solution phase space. In Section \ref{sec:on-shell-symplectic-from}, we {delve into the analysis of the}  on-shell symplectic form and compute Poisson brackets over the solution space. In Section \ref{sec:soln-space-carrollian}, we explore the geometry of bulk solution space {and demonstrate that it exhibits} a Carrollian structure. In Section \ref{Sec:subsectors}, we study three {distinctive} sectors {within} the solution space. Section \ref{sec:discussion} is devoted to {providing an} outlook and concluding remarks. {Within various appendices, we have compiled essential background information to enhance the self-contained nature of this paper, as well as in-depth calculations.} 
In Appendix \ref{appen-A}, we review some basic features of light-front scalar field theory, {including} its symplectic form and on-shell Poisson brackets. {We establish the equivalence between on-shell Poisson brackets and Dirac brackets on the off-shell solution phase space.} In Appendix \ref{appen-carrollian}, we review basics of Carrollian geometry.  Appendix \ref{appen-sym-gen}  discusses boundary symmetry generators and corresponding field variations. In Appendix \ref{appen-charge}, we discuss analysis of boundary symmetry charges and their algebra. In Appendix \ref{appen-on-shell-symp-details}, we {provide } details {of} computation of on-shell symplectic form. In Appendix \ref{appen-inverting}, we provide {details} of inverting on-shell bulk symplectic form and {extracting} the corresponding Poisson brackets.  In Appendix \ref{appen-4d} we present specification of the general results of the main text to the physically significant   $D=4$ case.

\section{Einstein Gravity on a Spacetime with {a} Null boundary, a quick review}\label{sec2:soln-space-review}


{Considering}  a null surface at $r=0$, we {define} $r$ {as} a coordinate {to} measures deviation from the given null surface. {We}  denote {the} advanced time by $v$ and the transverse coordinates by $x^A$, where $A=1,\cdots,D-2$. The line-element in the Gaussian null-type coordinate system takes the form
\begin{equation}\label{metric'}
    \d s^2= g_{\mu\nu}\d x^\mu \d x^\nu= -V \d v^2 + 2 \eta \d v \d r + q_{AB} \left( \d x^A + U^A \d v \right) \left( \d x^B + U^B \d v \right)\, ,
\end{equation}
where $x^\mu=\{v, r, x^A \}$.

\paragraph{Expanding around null surface.}
Here, $V, q_{AB}, U^A$ {represent }  unknown functions that have $r$ dependence, while $\eta$ does not depend on $r$. We {{\textit{presume}}} these {functions} are smooth and {amenable to} Taylor expansion  around $r=0$. The metric on transverse surface, $q_{AB}$ can be expanded as
\begin{equation}
    \begin{split}\label{metric-expansion'}
         q_{AB} = & \, \Omega_{AB} + 2\, r \,\eta \left( \lambda_{AB} + \frac{\lambda}{D-2} \, \Omega_{AB} \right) + \mathcal{O}(r^2)\, .
    \end{split}
\end{equation}
In our notation $\lambda_{AB}$ is traceless.
We also use the notation $\Omega:=\sqrt{\det \Omega_{AB} }\,$.  We can also expand $U^A$ and $V$ as
\begin{subequations}
    \begin{align}
       & U^A= {\cal U}^A - r \, \eta\, \Omega^{-1} {\cal J}^A + \mathcal{O}(r^2)\, , \\
        &V=-\eta \left(  {\Gamma - \frac{2}{D-2} \frac{\mathcal{D}_v \Omega}{\Omega}}+ \frac{\mathcal{D}_v \eta}{\eta} \right)  r  + \mathcal{O}(r^2)\, ,
    \end{align}
\end{subequations}
where ${\cal D}_v=\partial_{v}-\mathcal{L}_{\mathcal{U}}$, with ${\cal L}_{\mathcal{U}}$ is the Lie derivative along ${\mathcal{U}}$. 
\paragraph{Null geometric quantities.}
One may also define the induced metric {using} two null vector fields,
\begin{subequations}\label{l-n-vector}
    \begin{align}
       l_\mu \d x^\mu =-\frac{1}{2} V \d v + \eta \d r \, , & \qquad\qquad l^\mu \partial_\mu = \partial_v -U^A\partial_A + \frac{V}{2\eta} \partial_r\, ,  \\ 
         n_\mu \d x^\mu = - \d v \, , & \qquad\qquad n ^\mu \partial_\mu = - \frac{1}{\eta} \partial_r\, ,
    \end{align}
\end{subequations}
where $l^\mu$ and $ n^\mu$ are outward-pointing and inward-pointing, respectively, and they are normalized such that  $l\cdot n=-1$. The induced metric on the transverse surface can be {expressed} as $q_{\mu \nu}= g_{\mu \nu}+2 \,  l_{(\mu} n_{\nu)} $. The deviation tensor associated with {the} vector field $l^\mu$ is
\begin{equation}\label{}
       B _{\mu \nu} :=  \left(q^{\alpha}_{\mu} q^{\beta}_{\nu}\nabla_{\beta} l_{\alpha}\right)_{r=0} = \frac{1}{D-2}   \theta \, q_{\mu \nu} + N_{\mu \nu}\, , 
\end{equation}
where
\begin{equation}
{\theta} =\frac{\mathcal{D}_v \Omega}{\Omega} \, , \qquad  {N}_{AB}= \frac{1}{2} \mathcal{D}_v \Omega_{AB} -\frac{1}{D-2} \, \frac{\mathcal{D}_v\Omega}{\Omega}\, \Omega_{AB} \, ,
\end{equation}
{are respectively expansion and shear/news tensors.}
For later convenience, we introduce the following notations: 
\begin{subequations}\label{gamma-P-def}
\begin{align}
{\cal P}&:=\ln{\left(\eta\, \theta^{-2}\right)}\, ,\\
    \gamma_{{AB}} &:=\Omega^{-2/(D-2)}\Omega_{AB}\, ,\qquad \det{\gamma_{AB}}=1\, ,\\
   \mathcal{N}_{AB}&:=\frac{1}{2}\, \mathcal{D}_v \gamma_{AB}, \qquad N_{AB}=\Omega^{2/(D-2)} \mathcal{N}_{AB}\, .
\end{align}
\end{subequations}
We also assume {that} the unimodular metric $\gamma_{AB}$ is invertible and denote its inverse {as} $\gamma^{AB}$, i.e. 
$\gamma_{AB}\gamma^{BC}=\delta_A{}^C$.  As the notation {suggests}, these are scalar, vector, and tensor with respect to the codimension 2 diffeomorphisms along a constant $v$ slice of the null hypersurface located at $r=0$. 
These quantities {carry} 
$A,B$-type indices and we will use $\gamma_{AB}$ or $\gamma^{AB}$ to lower or raise indices on other quantities. 
\paragraph{Equations of motion and solution space.}

To construct the solution space, we need to impose Einstein's equations $G_{\mu\nu} + \Lambda g_{\mu \nu}=0$,
where $\Lambda$ {represents the} cosmological constant and $G_{\mu\nu}$ is the Einstein tensor. We can solve Einstein's equations {in an} order by order {manner} in powers of $r$. Among these equations, there are two conspicuous equations, {namely the} Raychaudhuri and Damour equations. Once we expand metric functions and parameters around $r=0$, we {acquire:} \cite{Adami:2021nnf,Adami:2021kvx}
\begin{itemize}
    \item  Two pairs of scalars, $(\Gamma , \Omega)$,
 $( \theta, {\cal P})$;
    \item A pair of vectors, $({\cal U}^A, \mathcal{J}_A)$;
    \item A pair of two-tensors, $({\cal N}_{AB}, \gamma^{AB})$;
\end{itemize}
These pairs are subject to equations of motion (constraints),\footnote{In writing {the} Raychaudhuri  \eqref{Raych} and Damour \eqref{Damour} equations, we assume {that} $\theta\neq 0$ (as we have divided these equations by $\theta$ to obtain these forms). {In} the special case, {where}  $\theta=0$, {the} Raychaudhuri equation leads ${\cal N}_{AB}=0$ and {the} Damour equation takes the form  $ {\cal D}_v{\cal J}_A+\Omega\partial_A\Gamma=0$ \cite{Adami:2021nnf}.} 
\begin{subequations}\label{4-EoM}
    \begin{align}
        & -\partial_v \Omega+ \vartheta  \approx 0\, , \label{theta-Omega}\\
        & -\partial_v  {\cal P}+ \varpi +  2 \, \theta^{-1} \, \mathcal{N}^{\,2} \approx 0\, ,\label{Raych}\\
        & -\partial_v  {\cal S}_A +{\partial}_{B}(\mathcal{U}^{B}\mathcal{S}_{A})+\mathcal{S}_{B}{\partial}_{A}\mathcal{U}^B+ 2 \, \Omega \,  \partial_A (\theta^{-1}\mathcal{N}^{\, 2})- 2 \nabla^B (\Omega \, \mathcal{N}_{AB}) \approx 0\, ,\label{Damour}\\
        &-\partial_v  {\gamma}_{AB} +\nabla_{A}\mathcal{U}_{B}+\nabla_{B}\mathcal{U}_{A}-\frac{2}{D-2}\nabla_{C}\mathcal{U}^{C}\gamma_{AB}
        + 2 \, \mathcal{N}_{AB} \approx 0\, , \label{News-News}
    \end{align}
\end{subequations}
where $\nabla_A$ {denotes} the covariant derivative w.r.t. unimodular metric $\gamma_{AB}$ and new quantities are defined as {follows:} 
\begin{equation}\label{S-varpi-vartheta}
    \mathcal{S}_A:={\cal J}_{A}+\Omega \partial_A{\cal P}\, , \qquad \varpi := \Gamma+\mathcal{U}^{A}\partial_{A}\mathcal{P},\qquad \vartheta:=\Omega\theta+\partial_A(\Omega{\cal U}^A)\, , \qquad \mathcal{N}^{\,2}:=\mathcal{N}_{AB}\mathcal{N}^{AB}\, .
\end{equation}
{$\mathcal{S}_A$} is the ``spin superrotations'' part (or superspin, for short) of the angular momentum (superrotations). {It is derived by}  subtracting the orbital superrotations $-\Omega\partial_A {\cal P}$ {from} the total superrotations ${\cal J}_A$.

Equations \eqref{theta-Omega} and \eqref{News-News} can be {regarded} as two equations of motion {that} specify {the} expansion and news tensor. The Damour \eqref{Damour} specifies {angular velocity aspect} $\mathcal{U^A}$ and the Raychaudhuri equation \eqref{Raych}  can be seen as {an} equation for $\varpi$. Therefore, the solution phase space for a null boundary is {entirely} {characterized} by two scalars $\Omega, {\cal P}$, one vector $\mathcal{S}_A$, and one tensor mode $\gamma_{AB}$ ({along with} their canonical conjugates).

\section{Off-shell Null Boundary Symplectic Form}\label{sec:symplectic-form}
For later convenience, we {adopt} the following notation:
\begin{equation}
    \oint_{\mathcal{N}_{b}} =  \int \d{}^{D-2}x \, , \qquad  \int_\mathcal{N} = \int \d v \int \d{}^{D-2}x\, .
\end{equation}
where ${\cal N}_b$ denotes a constant $v=v_b$ slice on the null boundary.  For simplicity, we also assume {that} the codimension 2 transverse surface ${\cal N}_b$ is compact.

\paragraph{Off-shell symplectic form.}
The {standard} Lee-Wald off-shell symplectic form can be read from the Einstein-Hilbert action as \cite{Adami:2021kvx}
\begin{equation}\label{off-shell-symplectic}
    \begin{split} 
     \bO_{\text{\tiny LW}}[\delta \Phi , \delta \Phi ; g]
        &=  \frac{1}{16 \pi G} \int_{\mathcal{N}} \bigr[ \, \delta \mathcal{U}^{A} \wedge\delta {\cal J}_{A}  - \delta \Gamma\wedge \delta \Omega 
        +\delta ( \Omega \, \theta ) \wedge \delta \mathcal{P}+ \delta {\cal N}_{AB}\wedge \delta (\Omega \, \gamma^{AB}) \, \bigr]\\ 
        &=  \frac{1}{16 \pi G} \int_{\mathcal{N}} \bigr[ \, \delta \mathcal{U}^{A} \wedge\delta {\cal S}_{A}  + \delta \Omega \wedge \delta \varpi
        +\delta \vartheta \wedge \delta \mathcal{P}+ \delta \mathcal{N}_{AB}\wedge \delta (\Omega \, \gamma^{AB}) \, \bigr].
    \end{split}
\end{equation}
where the above is subject to second-class constraints,
\begin{equation}\label{C1-C2}
    C_1=\det{\gamma}-1=0,\qquad C_2:=\gamma^{AB}{\cal N}_{AB}=0.
\end{equation}
In our adopted coordinate system, $r$ is taken to be the affine parameter along null geodesics generated by the null vector field $n^\mu$. {Hence,} it {bears a} {resemblance} to the usual time coordinate, {particularly} when we are consider {the} Hamiltonian formulation  based on {the} ADM decomposition. Strictly speaking, the bulk evolution is {orchestrated } by $r$. {Consequently},  the Lee-Wald symplectic form \eqref{off-shell-symplectic} can be {viewed} as a symplectic form written on a constant affine parameter.  Advanced time is one of the coordinates on the {specified}  hypersurface. 

{By} inverting the above symplectic form, one can read nonvanishing ``off-shell'' Poisson brackets
\begin{subequations}\label{off-shell-Poisson-brackets}
    \begin{align}
        & \left\{ \mathcal{U}^A(v_{\text{\tiny 1}}, \mathbf{x}_{\text{\tiny 1}}) , \mathcal{S}_B(v_{\text{\tiny 2}}, \mathbf{x}_{\text{\tiny 2}})\right\}_{\text{\tiny PB}}= 16 \pi G \, \delta^A_B \, \delta (v_{\text{\tiny 1}}-v_{\text{\tiny 2}}) \, \delta^{D-2}(\mathbf{x}_{\text{\tiny 1}}-\mathbf{x}_{\text{\tiny 2}})\, , \label{PB-a}\\
        &  \left\{ \Omega(v_{\text{\tiny 1}}, \mathbf{x}_{\text{\tiny 1}}) , \varpi(v_{\text{\tiny 2}}, \mathbf{x}_{\text{\tiny 2}})\right\}_{\text{\tiny PB}}= 16 \pi G  \, \delta (v_{\text{\tiny 1}}-v_{\text{\tiny 2}}) \, \delta^{D-2}(\mathbf{x}_{\text{\tiny 1}}-\mathbf{x}_{\text{\tiny 2}})\, , \label{PB-b}\\
        &  \left\{ \vartheta(v_{\text{\tiny 1}}, \mathbf{x}_{\text{\tiny 1}}) , \mathcal{P}(v_{\text{\tiny 2}}, \mathbf{x}_{\text{\tiny 2}})\right\}_{\text{\tiny PB}}= 16 \pi G  \, \delta (v_{\text{\tiny 1}}-v_{\text{\tiny 2}}) \, \delta^{D-2}(\mathbf{x}_{\text{\tiny 1}}-\mathbf{x}_{\text{\tiny 2}})\, , \label{PB-c}\\
        & \left\{ \mathcal{N}_{AB}(v_{\text{\tiny 1}}, \mathbf{x}_{\text{\tiny 1}}) , \Omega(v_{\text{\tiny 2}}, \mathbf{x}_{\text{\tiny 2}}) \, \gamma^{CD}(v_{\text{\tiny 2}}, \mathbf{x}_{\text{\tiny 2}})\right\}_{\text{\tiny DB}}= 16 \pi G \, P^{CD}_{AB}(v_{\text{\tiny 2}}, \mathbf{x}_{\text{\tiny 2}})\, \delta (v_{\text{\tiny 1}}-v_{\text{\tiny 2}}) \, \delta^{D-2}(\mathbf{x}_{\text{\tiny 1}}-\mathbf{x}_{\text{\tiny 2}})\, . \label{PB-N}
    \end{align}
\end{subequations}
where
\begin{equation}\label{P-matrix-def}
    P^{CD}_{AB}(v):=\delta^C_{(A} \delta^D_{B)} -\frac{1}{D-2}\gamma_{AB}(v)\gamma^{CD}(v)\, .
\end{equation}
is a projection operator, projecting tensors to their trace-less parts. 
In the {expression} above, $\delta^{D-2}(\mathbf{x}_{\text{\tiny 1}}-\mathbf{x}_{\text{\tiny 2}})$ is  {merely a} product of $D-2$ delta-functions and does not involve $\Omega$. {Additionally,} \eqref{PB-N} is the Dirac bracket, {which is} subjected to the constraints \eqref{C1-C2}, see appendix \ref{appen-eq(3-5)} for more details of the calculation. $P^{CD}_{AB}$ is symmetric w.r.t. $A,B$ and $C,D$ indices and it is traceless, $ P^{CD}_{AB} \gamma_{CD}= P^{CD}_{AB}\gamma^{AB}=0$. 
{It is worth noting} that \eqref{PB-N} {can} also {be expressed}  as 
\begin{equation}\label{PB-NN''}
    \left\{ \mathcal{N}_{AB}(x_{_1}) ,  \gamma_{CD}(x_{_2})\right\}_{\text{\tiny DB}}= -\frac{16\pi G}{ \Omega(x_{_2}) } {\cal G}_{ABCD}(x_{_2})
    \, \, \delta (v_{\text{\tiny 1}}-v_{\text{\tiny 2}}) \, \delta^{D-2}(\mathbf{x}_{\text{\tiny 1}}-\mathbf{x}_{\text{\tiny 2}})\, .
\end{equation}
where 
\begin{equation}\label{WdW-metric-def}
    {\cal G}_{ABCD}:=\frac12\left(\gamma_{AC}\gamma_{BD}+\gamma_{AD}\gamma_{BC}-\frac{2}{D-2}\gamma_{AB}\gamma_{CD}\right)\, ,
\end{equation}
is the Wheeler-DeWitt (WdW) metric \cite{DeWitt:1967yk}. This {exhibits} the following symmetry structures and  trace properties,
\begin{equation}\label{WdW-sym-trace}
\begin{split}
    {\cal G}_{ABCD}={\cal G}_{BACD}=&{\cal G}_{ABDC}={\cal G}_{CDAB}\, ,\\
\gamma^{AB}{\cal G}_{ABCD}=\gamma^{CD}{\cal G}_{ABCD}=0\, , &\qquad \gamma^{AC}{\cal G}_{ABCD}= \frac{D(D-3)}{2(D-2)}\gamma_{BD}\, .
\end{split}
\end{equation}
Note that the above Poisson brackets are still subject to the ``constraint equations'' 
\eqref{4-EoM}. We will impose  these constraints in the next section.

\section{On-shell Symplectic Form and Dirac Brackets}\label{sec:on-shell-symplectic-from}

One {can}  view the equations of motion \eqref{4-EoM} as (second-class) constraints and compute Dirac brackets. Alternatively, {these equations can be}  directly inserted  into the symplectic form \eqref{off-shell-symplectic} {to} compute the ``on-shell symplectic form''. In the {conventional} constrained system terminology, this {process involves } solving the constraints and going to the reduced phase space. We have shown in appendix \ref{appen-A} that, for a null-front field theory, these two methods yield the same result. {In the other words}, {the} Poisson brackets {derived } from the ``on-shell'' symplectic form defined on the reduced phase space (solution space) and the Dirac brackets using half of the Hamilton's equations as constraints (over the solution phase space) are {equivalent}. 

In our {particular} case, on the reduced phase space, we have one canonical pair of scalars, one vector, and one tensor mode. We {opt} to solve \eqref{Raych} for $\varpi$ (while keeping ${\cal P}$), {solve}  \eqref{theta-Omega} for $\vartheta$ (while {keeping}  $\Omega$), solve \eqref{Damour} for ${\cal U}^A$ (while keeping the spin superrotations ${\cal S}_A$) and {ultimately } {solve} \eqref{News-News} {to determine} ${\cal N}^{AB}$ (while {keeping}  $\gamma_{AB}$). Explicitly, 
\begin{subequations}\label{Gamma-theta-U-solved-for}
    \begin{align}
    & \vartheta={\partial_v \Omega}\, , \qquad {\varpi}={\partial}_v  {\cal P} -\frac{2}{\theta}\,  \mathcal{N}^{\,2}\, , \qquad  \mathcal{N}_{AB} =\frac12\mathcal{D}_v \gamma_{AB}  \, , \\
    & {\cal U}^A=  {\bar{\cal U}}^A ({\cal S}_A)+ ({\cal O}^{-1})^{AB} \left(\partial_v{\cal S}_B-2 {\cal X}_B\right)\, , \quad {\cal O}_{AB}:= {\cal S}_A\partial_B+{\cal S}_B\partial_A+\partial_B{\cal S}_A\, ,
        \label{calU-EoM}
    \end{align}
\end{subequations}
where  ${\cal O}_{AB} {\bar{\cal U}}^B=0$, ${\cal X}_A=  \partial_A (\theta^{-1}\mathcal{N}^{\, 2})- \nabla^B (\Omega \, \mathcal{N}_{AB})$ and 
$({\cal O}^{-1})^{AB}$ is the inverse of ${\cal O}_{AB}$ which has a linear functional dependence on ${\cal S}_A$. 
\subsection{Vanishing News \texorpdfstring{${\cal N}_{AB}=0$}{} case}
For this case, {by plugging}  \eqref{Gamma-theta-U-solved-for} into the symplectic form, we {obtain} \cite{Adami:2021kvx}
\begin{equation}\label{Omega-bdry}
   \begin{split} 
 \bO_{\text{\tiny LW}} &=  \frac{1}{16 \pi G} \int_{\mathcal{N}}  \ \partial_v\bigl(\delta \Omega \wedge \delta \mathcal{P}+ {\cal F}^{AB}\delta \mathcal{S}_{A} \wedge\delta {\cal S}_{B}\bigr) 
 \\ &=\frac{1}{16 \pi G} \oint_{\mathcal{N}_b}  \bigl[\delta \Omega \wedge \delta \mathcal{P}+ {\cal F}^{AB}\delta \mathcal{S}_{A} \wedge\delta {\cal S}_{B}\bigr]\, , 
 \end{split}
\end{equation}
where ${\cal F}^{AB}=(\{{\cal S}_A, {\cal S}_B\})^{-1}$ (see below). { It is evident that}  in the absence of ${\cal N}_{AB}$ modes, {the} on-shell symplectic form {simplifies into}  a boundary term {represented as an} integral over ${\cal N}_b$ \cite{Sheikh-Jabbari:2022mqi}.  That is, when ${\cal N}_{AB}=0$ the system is {solely} described  by boundary dof. {The} symplectic form \eqref{Omega-bdry} is a closed 2-form and is invertible. The inverse symplectic form yields Poisson brackets, 
\begin{equation}\label{bdry-brackets-1}
\begin{split}
    \{\Omega (v_b,\mathbf{x}), {\cal P}({v}_b,\mathbf{y})\}_{\text{\tiny{DB}}}&=16\pi G \  \delta^{D-2}(\mathbf{x}-\mathbf{y})\, ,\\
    \{\Omega (v_b,\mathbf{x}), \Omega(v_b,\mathbf{y})\}_{\text{\tiny{DB}}}&=\{ {\cal P}(v_b,\mathbf{x}), {\cal P}(v_b,\mathbf{y})\}_{\text{\tiny{DB}}}=0\, ,
\end{split}\end{equation}
\begin{equation}\label{bdry-brackets-2}
\begin{split}
    \{\Omega (v_b,\mathbf{x}), {\cal S}_A(v_b,\mathbf{y})\}_{\text{\tiny{DB}}}&= \{{\cal P}(v_b,\mathbf{x}), {\cal S}_A(v_b,\mathbf{y})\}_{\text{\tiny{DB}}}=0\, ,\\
    \{{\cal S}_A (v_b,\mathbf{x}), {\cal S}_B(v_b,\mathbf{y})\}_{\text{\tiny{DB}}}&=16\pi G\ \left({\cal S}_{A}(v_b,\mathbf{y})\frac{\partial}{\partial x^B}-{\cal S}_{B}(v_b,\mathbf{x})\frac{\partial}{\partial y^A}\right) \delta^{D-2}(\mathbf{x}-\mathbf{y})\, .
\end{split}
\end{equation}
The above is Heisenberg $\oplus$ Diff(${\cal N}_b$) algebra discussed in \cite{Adami:2021kvx}, see also \cite{Taghiloo:2022kmh, Sheikh-Jabbari:2022mqi}.

\subsection{Nonvanishing \texorpdfstring{${\cal N}_{AB}$}{} and Co-rotating Case}\label{sec:corotating}
To keep equations less cumbersome and to illustrate how turning on the news affects the symplectic form, we study the co-rotating case with ${\cal U}_A=0$. In this case, the on-shell symplectic potential and symplectic form respectively take the form
\begin{subequations}
    \begin{align}
        & 16\pi G \ \bTh_{\text{\tiny on-shell}} =  \oint_{\mathcal{N}_b}\, \Omega \, \delta \mathcal{P}    - \int_{\mathcal{N}}  \, \Omega \mathcal{N}^{AB}\, \left[   \delta  \gamma_{AB} -2 \, \frac{{\cal N}_{AB}}{\Omega \theta} \, \delta \Omega \right]\, , \label{Theta-on-shell}\\ 
        & 16\pi G \ \bO_{\text{\tiny on-shell}} =  \oint_{\mathcal{N}_b} \delta \Omega \wedge \delta \mathcal{P}    + \int_{\mathcal{N}} \biggl[- \delta \Omega^2 \wedge \delta \left(\frac{{\cal N}^2}{\partial_v\Omega}\right)        + \frac12\partial_v\delta \gamma_{AB}\wedge \delta (\Omega \, \gamma^{AB}) \biggr]\, .\label{Omega-on-shell}
    \end{align}
\end{subequations}
In the above, ``on-shell'' means imposing \eqref{4-EoM} with  $\mathcal{U}^A=0$, explicitly, this means
\begin{equation}\label{contraints-new'}
     \theta  =\frac{\partial_v \Omega}{\Omega}\, , \qquad 
    \mathcal{N}_{AB}=\frac12\partial_v  {\gamma}_{AB}\, .
\end{equation}
The Damour equation takes the form $   \partial_v  {\cal S}_A = 2 \, \Omega \,  \partial_A (\theta^{-1}\mathcal{N}^{\, 2})- 2 \nabla^B (\Omega \,  {\cal N}_{AB})$ which fully specifies the time dependence of the angular momentum aspect. {Strictly speaking, the Damour equation specifies $\mathcal{S}_A$ up to codimension 2 functions, {denoted as} $\bar{\mathcal{S}}_A(v_b,\boldsymbol{x})$. These integration constants {provide} us {with} angular momentum charges.} {Therefore}, the co-rotating reduced phase space is spanned by $\Omega\, , \mathcal{P} \, , \gamma_{AB}$.

The on-shell symplectic form consists of a codimension 1 integral over ${\cal N}$ and a codimension 2 part, integral over ${\cal N}_b$. However, there still {remains an}  arbitrariness in separating them into boundary and bulk parts, as $\oint_{{\cal N}_b} X=\int_{{\cal N}} \partial_v X$. This arbitrariness may be removed upon some other (physical) requirements. For example, we may require that the boundary and bulk parts are closed 2-forms over the solution space:
\begin{subequations}
\begin{align} 
&\bO_{\text{\tiny on-shell}} =   \bO^c_{\text{\tiny bdy}} + \bO^c_{\text{\tiny Bulk}}\, ,\\  
\bO^c_{\text{\tiny bdy}} =\frac{1}{16\pi G}\oint_{\mathcal{N}_b} \delta \Omega \wedge \delta \mathcal{P}\ ,  \qquad 
\bO^c_{\text{\tiny Bulk}}&=\frac{1}{16\pi G}\int_{\mathcal{N}} \biggl[- \delta \Omega^2 \wedge \delta \left(\frac{{\cal N}^2}{\partial_v\Omega}\right)    + \frac12\partial_v\delta \gamma_{AB}\wedge \delta (\Omega \, \gamma^{AB}) \biggr]\, . \label{Omega-on-shell-closed}
    \end{align}
\end{subequations}

As another way to separate the symplectic form into bulk and boundary parts {is to stipulate} that the bulk part {follows} the generic form {outlined} in appendix \ref{appen-A}, i.e. the bulk part {takes }  the form {$\int_{{\cal N}} \slashed{\delta} X \wedge \partial_v \slashed{\delta} X$, where $\slashed{\delta}$ is used to emphasis that $\slashed{\delta}X$ is not necessarily a closed 1-form on the phase space.} 
After some manipulations, the details of which are given in appendix \ref{appen-on-shell-symp-details}, one arrives at {the following expression:}
\begin{subequations}\label{main-result}
\begin{align} 
    \bO_{\text{\tiny on-shell}}&= \bO_{\text{\tiny bdy}}+ \bO_{\text{\tiny Bulk}}\, ,\\
\bO_{\text{\tiny bdy}} = \frac{1}{16\pi G}\oint_{\mathcal{N}_b} \delta \Omega \wedge \slashed{\delta}\hat{{\cal P}} \, ,
 \qquad & \bO_{\text{\tiny Bulk}}= \frac{1}{32\pi G}\int_{\mathcal{N}} \Omega {\cal G}^{ABCD}\ \slashed{\delta}{\hat{\gamma}}_{AB}\wedge \partial_v \slashed{\delta}{\hat{\gamma}}_{CD}\, , \label{Omega-bdry-bulk}
    \end{align}
\end{subequations}
where non-closed 1-forms  are defined as 
\begin{subequations}\label{delta-slashes}
\begin{align}
\slashed{\delta}\hat{{\cal P}}:=&\delta \mathcal{P}-  \frac{{\cal N}^{AB}}{\theta} \delta\gamma_{AB}= \delta \mathcal{P}+ \frac{\Omega\partial_v\gamma^{AB}}{2\partial_v\Omega} \delta\gamma_{AB}\, ,\label{slashed-delta-P}\\
\slashed{\delta}{\hat{\gamma}}_{AB}:=&\delta\gamma_{AB}-\frac{2{\cal N}_{AB}}{\Omega\theta}\delta\Omega =\delta\gamma_{AB}-\frac{\partial_v\gamma_{AB}}{\partial_v\Omega} \delta\Omega\, ,\label{slashed-delta-gamma}
\end{align}\end{subequations} 
and
\begin{equation}\label{WD-metric}
    \mathcal{G}^{ABCD}:=\frac{1}{2} \left(\gamma^{AC}\gamma^{BD}+ \gamma^{AD}\gamma^{BC}- \frac{2}{D-2} \gamma^{AB} \gamma^{CD}\right)\, ,
\end{equation}
is the inverse WdW metric \eqref{WdW-metric-def} which { possesses}  the same properties  in \eqref{WdW-sym-trace} and
\begin{equation}\label{WD-metric-inverse}
    \mathcal{G}^{ABEF}\mathcal{G}_{EFCD}=P^{AB}_{CD},\qquad 
    P^{AB}_{CD}\mathcal{G}^{CDEF}=\mathcal{G}^{ABEF}, \qquad P^{AB}_{CD}\mathcal{G}_{ABEF}=\mathcal{G}_{CDEF}.
    \end{equation}

The relation between the  two boundary/bulk separations {mentioned above} is as follows:
{\begin{equation}\label{closed-Omega-F-bdy}
      \bO_{\text{\tiny bdy}} = \bO^{c}_{\text{\tiny bdy}}
      + \mathbf{F}_{\text{\tiny bdy}}, \qquad \bO_{\text{\tiny Bulk}} = \bO^{c}_{\text{\tiny Bulk}}- \mathbf{F}_{\text{\tiny bdy}}\, ,
\end{equation}
where
\begin{equation}\label{F-bdry}
    \mathbf{F}_{\text{\tiny bdy}} = -\frac{1}{16\pi G} \oint_{\mathcal{N}_b}   \frac{{\cal N}^{AB}}{\theta} \delta \Omega \wedge \delta\gamma_{AB}=  -\frac{1}{16\pi G} \oint_{\mathcal{N}_b}   \frac{{\cal N}^{AB}}{\theta} \delta \Omega \wedge \slashed{\delta}{\hat{\gamma}}_{AB}\,,
\end{equation}
{represents the} boundary symplectic flux due to {the} passage of the bulk flux. $\bO^{c}_{\text{\tiny bdy}}$ {as} given in \eqref{Omega-bdry-bulk} is closed and invertible over ${\cal N}_{b}$, {yielding} the boundary Poisson brackets \eqref{bdry-brackets-1}. {On the other hand,} $\bO^{c}_{\text{\tiny Bulk}}$, while closed,  is not invertible over ${\cal N}$. We will discuss this further in the next sections.

\paragraph{Further comments.} Given the above on-shell symplectic forms, some comments are in order:

\begin{enumerate}[label=\Roman*.]
    \item {The} appearance of non-exact solution space one-forms $\slashed{\delta}\hat{{\cal P}}$ and $ \slashed{\delta}{\hat{\gamma}}_{AB}$ is the hallmark of the fact that neither of the boundary nor bulk parts of the phase space are individually closed. 
    \item While the boundary and bulk terms are not closed, the {combined } bulk+boundary system is closed, $\delta\bO_{\text{\tiny on-shell}} =0$.
\item {The closure }   of {the} bulk+boundary symplectic form is  necessary, but not {a} sufficient condition for  invertibility of the symplectic form. {Upon} closer inspection, {it becomes apparent}  that $\bO_{\text{\tiny on-shell}}$ is not invertible over the solution space on ${\cal N}$.

\item The non-closedness of individual bulk and boundary parts is a manifestation of the non-integrability of charges associated with symmetry generators, {as} discussed in appendix \ref{appen-sym-gen}. See also appendix \ref{appen-charge} for the analysis of surface charges and their algebra.

\item Non-closed parts of $\slashed{\delta}\hat{{\cal P}}$ and $ \slashed{\delta}{\hat{\gamma}}_{AB}$ are proportional to the news ${\cal N}_{AB}$.  {When} $\slashed{\delta}{\hat{\gamma}}_{AB}=0$, {it indicates}  a degeneracy direction of the bulk symplectic form $\bO_{\text{\tiny Bulk}}$. Nonetheless, {note that} $\bO_{\text{\tiny Bulk}}$ is invertible in the (codimension 1) subspace spanned by $\gamma_{AB}$. {In the next section,} we will show  that the bulk part of the on-shell solution space is indeed a Carrollian geometry. {By} inverting the invertible parts of the bulk and boundary symplectic forms, namely \label{Item-V}
\begin{equation}\label{Integ-Omega}
\bO^{\text{\tiny{I}}}_{\text{\tiny Bulk}}={\frac{1}{32 \pi G}}\int_{\mathcal{N}} \Omega\, {\cal G}^{ABCD}\ {\delta}{{\gamma}}_{AB}\wedge \partial_v {\delta}{{\gamma}}_{CD}\, , \hspace{1 cm} \bO^{\text{\tiny{I}}}_{\text{\tiny bdy}} = {\frac{1}{16 \pi G}}\oint_{\mathcal{N}_b} \delta \Omega \wedge \delta {\cal P}\, ,
\end{equation}
over the null boundary ${\cal N}$ {and ${\cal N}_{b}$ respectively} yields (see appendix \ref{appen-bulk-on-shell} for more details)
\begin{subequations}
        \begin{align}  
        { \{\Omega (v_b,\mathbf{x}), \mathcal{P}(v_b,\mathbf{x}')\}}& = 16 \pi G \, \delta^{D-2}(\mathbf{x}-\mathbf{x}')\, ,\\
            \{\Omega (v_b,\mathbf{x}), \gamma_{AB}(v,\mathbf{x}')\}&= 0\, , \\ 
               \{{\cal P} (v_b,\mathbf{x}),  \gamma_{AB}(v,\mathbf{x}')\}&= 0\ , \\
  \{\gamma_{AB}(v,\mathbf{x}), {\gamma}_{CD}(v',\mathbf{x}')\}&= \frac{{16\pi G}}{\sqrt{\Omega(x)\Omega(x')}}    \,  \bohm_{ABCD}^{\text{\tiny{S}}}(v,v';\mathbf{x},\mathbf{x}') \, H(v-v')\ \delta^{D-2}(\mathbf{x}-\mathbf{x}')\, , \label{Bulk-PB} 
               \end{align}
    \end{subequations}
{where $\bohm^{\text{\tiny{S}}}_{ABCD}(v,v';\mathbf{x},\mathbf{x}')$ is given by \eqref{bohm-vv'-final}.}

{Notice that to obtain \eqref{Bulk-PB} we assume $D>3$.} For $D=3$ pure Einstein gravity, we do not have any bulk modes, explicitly $\gamma_{AB}$. In $D=3$ case the solution space is {solely} governed by the boundary modes. See \cite{Adami:2020ugu} for a detailed analysis.

\end{enumerate}

\section{Bulk Part of Solution Phase Space is a Carrollian Geometry}\label{sec:soln-space-carrollian}

Recalling the analysis of appendix \ref{appen-A}, in particular, \eqref{on-shell-sym-form-scalar}, the on-shell symplectic form is expected to carry information about the metric over the field space. Let us focus on the bulk part of the symplectic form
\begin{equation}\label{Omega-Metric}
    {\bO_{\text{\tiny Bulk}}} =  \frac{1}{32\pi G} \, \int_{\mathcal{N}}\, \int^{\prime}_{\mathcal{N}} \, \delta \varphi^{\mathbb{I}}(x) \, \bO_{_{\mathbb{IJ}}}[x;x']\wedge \delta \varphi^{\mathbb{J}}(x')\, ,
\end{equation}
where $\varphi^{\mathbb{I}}=\{   \Omega \, , \mathcal{\gamma}_{AB}\}$ denotes the bulk phase space variables and  $\bO_{_{\mathbb{IJ}}}[x;x']$ is a $(D-1)(D-2)/2$ dimensional antisymmetric matrix. {Recalling \eqref{Omega-bdry-bulk}}
we can write it as
\begin{equation}
    \bO_{_{\mathbb{IJ}}}[x;x']=\bGIJ [x;x'] \,\partial_v\delta(v-v') \, \delta^{D-2}(\mathbf{x}-\mathbf{x}')\, ,
\end{equation}
where $\bGIJ [x;x']=\bGJI [x';x]$ and
\begin{equation}\label{bulk-soln-metric}
\bGIJ [x;x']=\begin{pmatrix}
 {\bG}_{_{\Omega\Omega  }}[x;x'] & ({\bG}_{_{\Omega\gamma}})^{AB}[x;x'] \\
 ({\bG}_{_{\gamma\Omega}})^{AB}[x;x'] &  (\bG_{_{\gamma \gamma}})^{ABCD}[x;x']
\end{pmatrix}  \, ,
\end{equation}
with 
\begin{subequations}\label{bulk-soln-metric-compt}
    \begin{align}
        &{\bG}_{_{\Omega\Omega  }}[x;x'] = 4\, ({\bG}_{_{\gamma \gamma}})^{ABCD}[x;x']\frac{{\cal N}_{AB}(x)}{\Omega(x)\theta (x)} \frac{{\cal N}_{CD}(x')}{\Omega(x')\theta (x')}  
        \, ,\\ 
        &({\bG}_{_{\Omega\gamma}})^{AB}[x;x'] = - 
        2\, ({\bG}_{_{\gamma \gamma}})^{ABCD}[x;x']\frac{{\cal N}_{CD}(x)}{\Omega(x)\theta (x) }\, ,\\
        & ({\bG}_{_{\gamma \gamma}})^{ABCD}[x;x'] =  \sqrt{\Omega(x)\Omega(x')}\,  {\cal G} ^{ABCD}[x;x']\,  ,
    \end{align}
\end{subequations}
where ${\cal G} ^{ABCD}[x;x']$ is the \textit{point-split WdW metric}: 
\begin{equation}\label{point-split-WdW}
     \begin{split}
         {\cal G} ^{ABCD}[x;x']=\frac14 \Bigl[ & \gamma^{AC}(x)\gamma^{BD}(x')+\gamma^{AD}(x)\gamma^{BC}(x')+\gamma^{AC}(x')\gamma^{BD}(x)\\
         &+\gamma^{AD}(x')\gamma^{BC}(x)-\frac{{4}}{D-2}\gamma^{AB}(x)\gamma^{CD}(x')\Bigr]\, .
     \end{split}
\end{equation}
such that ${\cal G} ^{ABCD}[x;x]=\mathcal{G}^{ABCD}(x)$ is the traceless WdW metric given in \eqref{WD-metric}.

The symmetric matrix $\bGIJ$ is the metric over the bulk part of the Lagrangian submanifold of the solution phase space. Explicitly, we may define ``line element'' over this space as:
\begin{equation}\label{Bulk-METRIC}
\begin{split}
 \delta{\mathbf{\mathbb{S}}}^2 &:= \int_{\cal N} \int^{\prime}_{\cal N} \bGIJ [x;x'] \delta\varphi^\mathbb{I}(x) \, \delta\varphi^\mathbb{J}(x')\, = \int_{\cal N} \int^{\prime}_{\cal N} \sqrt{\Omega(x)\Omega(x')}\,  {\cal G} ^{ABCD}[x;x']\slashed{\delta}{\hat\gamma}_{AB}(x)\slashed{\delta}{\hat\gamma}_{CD}(x')
\\ &=\int_{\cal N} \int^{\prime}_{\cal N}  \sqrt{\Omega(x)\Omega(x')}\,  {\cal G} ^{ABCD}[x;x']
         \left(\delta\gamma_{AB}(x)- \frac{\partial_v\gamma_{AB}(x)}{\partial_v\Omega(x)}\delta\Omega (x)\right)\left(\delta\gamma_{CD}(x')- \frac{\partial_{v'}\gamma_{CD}(x')}{\partial_{v'}\Omega(x')}\delta\Omega (x')\right) .
 \end{split}
\end{equation}
This manifold is describing a Carrollian geometry (see appendix \ref{appen-carrollian} for more discussions), as it has a  kernel vector
\begin{equation}
    K^{\mathbb{I}}[x]
    \, \frac{\delta}{\delta \varphi^\mathbb{I}}=\frac{\delta}{\delta \Omega} + \frac{2\mathcal{N}_{AB}}{\Omega \theta} \, \frac{\delta}{\delta \gamma_{AB}}\, ,
\end{equation}
such that $\bGIJ [x;x'] K^{\mathbb{J}}[x']=0$ and the $D(D-3)/2$ dimensional metric $\sqrt{\Omega(x)\Omega(x')}\,  {\cal G} ^{ABCD}[x;x']$. The integral curves of this  kernel vector on the phase space are given by 
\begin{equation}
    \gamma_{AB}(x)= \gamma_{AB}(\Omega(x),x^C)+ \tilde{\gamma}_{AB}(v_b,x^C)\, .
\end{equation}
We can also introduce an extra structure and define the Ehresmann connection 
\begin{equation}
    E_{\mathbb{I}} \, \delta \varphi^{\mathbb{I}}= \delta \Omega+ X^{AB} \slashed{\delta}{\hat{\gamma}}_{AB}\, , 
\end{equation}
such that $K^{\mathbb{I}}[x]E_{\mathbb{I}}[x]=1$. One can choose $X^{AB}$ to be zero, without loss of generality.
This Carrollian structure sheds further light on the bulk part of the symplectic form that is non-invertible.

\section{Various Sub-sectors of the Co-rotating Solution Space}\label{Sec:subsectors}
In this section, we {explore} the various sub-sectors of the solution space. The first {notable} sector is {characterized by } the vanishing news ${\cal N}_{AB}=0$ {within} the solution space.  In this {case,} boundary surface charges become integrable. In an appropriate slicing of the solution space, {this} yields a Heisenberg plus $\emph{Diff}\,({\cal N}_b)$ algebra \eqref{bdry-brackets-1}, \eqref{bdry-brackets-2}. This sector has been  {extensively studied} in previous works \cite{Adami:2021nnf, Adami:2021kvx, Sheikh-Jabbari:2022xix} and we {will} not repeat it {here}. In the rest of this section, we discuss other interesting subsectors.

\subsection{Outgoing Robinson-Trautman Gravitational Wave Sector} 
From \eqref{delta-slashes} we {infer} that when $\slashed{\delta}\hat{{\cal P}}=0$ or $\slashed{\delta}{\hat{\gamma}}_{AB}=0$, respectively the boundary or bulk directions of the symplectic form vanish. When $\slashed{\delta}{\hat{\gamma}}_{AB}=0$,
\begin{equation}\label{RT-01}
    \mathcal{N}_{AB}= \frac{1}{2} \, \frac{\partial \Omega}{\partial v } \, \frac{\delta \gamma_{AB}}{\delta \Omega } \qquad \Rightarrow \qquad \gamma_{AB}({x})= \gamma_{AB}(\Omega({x}),\mathbf{x})+ \tilde{\gamma}_{AB}(v_b,\mathbf{x})\, .
\end{equation}
That is, $\slashed{\delta}{\hat{\gamma}}_{AB}=0$ along the integral curves of the  kernel vector on the Carrollian solution space.
The above describes the (non-rotating) Robinson-Trautman ``spherical gravitational waves'' solutions \cite{Robinson:1960zzb, Robinson:1962zz}.  
One can readily check that $\mathbf{F}_{\text{\tiny bdy}}$ vanishes and hence $\bO_{\text{\tiny on-shell}} = \bO^{\text{\tiny{c}}}_{\text{\tiny bdy}}$.\footnote{In this sector 
$\slashed{\delta}\hat{{\cal P}}= {\delta}{\cal P} -\frac{2{\cal N}^2}{\Omega \theta^2}\delta\Omega={\delta}{\cal P}+ \delta {\cal F}(\Omega)$, 
where $\delta {\cal F}/\delta\Omega={\frac{1}{2}}\frac{\delta \gamma_{AB}}{\delta \Omega } \frac{\delta \gamma^{AB}}{\delta \Omega } \Omega$.}
This yields Poisson brackets \eqref{bdry-brackets-1}.

The Robison-Trautman solutions are particular gravitational waves in {several respects.} They are {entirely} {characterized} by the scalar (density) mode $\Omega$ {defined} over the null surface ${\cal N}$ {as opposed to}  the traceless-symmetric modes ${\cal N}_{AB}$. {These solutions} do not contribute to the bulk or boundary symplectic forms as well as the boundary symplectic flux, {given in} \eqref{closed-Omega-F-bdy}. In our analysis $\Omega$ {serves a dual} role: {it acts as } a boundary mode, associated with the entropy density in the null surface thermodynamics description \cite{Adami:2021kvx} and {as} a bulk mode, parameterizing the Robinson-Trautman solution.

\subsection{Non-expanding Null Boundaries}\label{sec:non-expanding}
In this subsection, we consider the non-expanding null surfaces $\theta=0$ which, using boundary equations of motion, yields $\Omega=\Omega({v_b},\mathbf{x})$. In this case, the Raychaudhuri equation {leads to}  $\mathcal{N}_{AB}=0$ \cite{Adami:2021nnf, Adami:2021kvx}. The Damour equation {simplifies to} 
${\cal D}_v \mathcal{J}_A+\Omega \partial_A \Gamma=0$.  For the ``co-rotating'' (${\cal U}^A=0$) case we are considering here, {this equation} reduces to $\partial_v \mathcal{J}_A=-\Omega  ({v_b},\mathbf{x}) \partial_A \Gamma$, {which} fixes {the} $v$-dependence of $\mathcal{J}_A$ in terms of $\Gamma=\Gamma(v,\mathbf{x})$. The on-shell symplectic form  in {the} non-expanding case becomes
\begin{equation}\label{presymplectic-NE-1}
    \begin{split} 
     \bO_{\text{\tiny NE}}[\delta \Omega , \delta \Gamma]
        =  \frac{1}{16 \pi G} \int_{\mathcal{N}}\delta \Omega({v_b},\mathbf{x}) \wedge  \delta \Gamma(v,\mathbf{x})\, .
    \end{split}
\end{equation}
If $\Gamma(v,\mathbf{x})=\partial_{v}\rho  (v,\mathbf{x})$, then $\mathcal{J}_A=-\Omega  ({v_b},\mathbf{x}) \partial_A\rho+ \bar{{\cal J}_A} (v_b,\mathbf{x})$ and the on-shell symplectic form takes the form
\begin{equation}\label{presymplectic-NE-2}
    \begin{split} 
     \bO_{\text{\tiny NE}}[\delta \Omega , \delta \rho]
        = \frac{1}{16 \pi G} \oint_{\mathcal{N}_{b}}\delta \Omega({v_b},\mathbf{x}) \wedge  \delta \rho(v_b, \mathbf{x})\, .
    \end{split}
\end{equation}
This yields the on-shell Poisson brackets 
\begin{equation}\label{NE-PB}
 \{\Omega ({v_b},\mathbf{x}), \Omega ({v_b},\mathbf{y})\}=0\, ,  \hspace{5 mm}  \{\Omega ({v_b},\mathbf{x}),\rho (v_b,\mathbf{y})\}=16 \pi G \delta^{D-2}(\mathbf{x}-\mathbf{y})\, ,   \hspace{5 mm}  \{\rho (v_b, \mathbf{x}), \rho (v_b,\mathbf{y})\}=0\, .
\end{equation}
So, we have a quantum mechanical system at ${\cal N}_b$ for the non-expanding non-rotating case. {Strictly speaking, we {have} turned off all of the gravitational waves, including outgoing Robinson-Trautman mode, and hence $\Omega$ becomes {solely a} boundary mode. }

\subsection{Decoupling of Bulk Modes from Boundary Modes}\label{sec:vanishing delta-omega}

As discussed, $\Omega$ is a special degree of freedom {with a dual role-} it appears {both} as a boundary and a bulk mode, and {it} couples the boundary and bulk parts of the symplectic form. {We can identify} a special subsector of the solution space, {by imposing the  $\delta \Omega=0$ condition}. As it is {evident} from \eqref{Omega-bdry-bulk} {that} when $\delta \Omega=0$ the boundary symplectic form vanishes, and the bulk part becomes {both} closed and invertible. {Specifically,} in this sector
\begin{equation}
    \bO_{\text{\tiny on-shell}}=\frac12\int_{\mathcal{N}} \Omega\, {\cal G}^{ABCD}\ {\delta}{{\gamma}}_{AB}\wedge \partial_v {\delta}{{\gamma}}_{CD}\, .
\end{equation}
This symplectic form is invertible and yields  Poisson brackets \eqref{Bulk-PB} (see appendix \ref{appen-inverting} for more details).

\section{Discussion and Outlook}\label{sec:discussion}

Building {upon the} results {presented} in \cite{Adami:2021nnf, Adami:2021kvx, Sheikh-Jabbari:2022mqi}, where {the} null boundary solution (phase) space for $D$ dimension was constructed, and surface charges analyzed, we focused more on the symplectic structure on the solution space and the {associated}  Poisson brackets. The motivation {behind} this study is paving the {way}  for the ``quantization'' of the null boundary solution space. This endeavor is aimed at addressing {fundamental} questions {in} black hole physics, {particularly} the microstates, and the information problem. {The null boundary} solution space {is intricately structured, comprising both} boundary and bulk parts that {interact}  through the symplectic flux \eqref{F-bdry}. For stationary black holes, we {deal with} the solution space restricted to $\theta=0$, cf. section \ref{sec:non-expanding}. In this case, the symplectic flux vanishes and the boundary system is described by $\Omega(v_b,\mathbf{x}), \rho(v_b,\mathbf{x})$ that form a Heisenberg algebra in which $16\pi G$ plays the role of $\hbar$. In this system, $\Omega$ is the entropy density, {while}  $\rho$ is its canonical/thermodynamical conjugate, playing the role of temperature. Assuming that ${\cal N}_b$ is compact {with} finite volume, upon {applying} semiclassical quantization, i.e. promoting the solution space to {a} Hilbert space, $\Omega$ and $\rho$ to operators {while substituting }  Poisson bracket \eqref{NE-PB} with commutators, we get quantization of the ``horizon area'' and/or variations {in} temperature. {Further investigation into this semiclassical quantization could provide valuable insights.} 

{Both} the bulk and boundary systems are open {as} {their symplectic 2-forms are nonclosed.} {However, when combined as} the bulk+boundary {system, the symplectic form becomes} closed. Nonetheless, the symplectic form of the whole system is not invertible, {indicating that the}  Poisson brackets in {the} invertible subsectors do not carry the whole information in the system. This {non-invertibility} has two sources: {the} existence of symplectic flux \eqref{F-bdry} and the Carrollian structure of the bulk solution space. This Carrollian structure {is} in the solution space and should not be mistaken with the Carrollian geometry of {the} null boundary ${\cal N}$. The former is infinite-dimensional, dealing with functions over ${\cal N}$, {and occurs in} the $(D-1)(D-2)/2$ dimensional matrices at any given point on ${\cal N}$. The null direction on the Carrollian solution space corresponds to the class of Robinson-Trautman (RT) gravitational waves \cite{Robinson:1960zzb, Robinson:1962zz}. {The existence}  of Carrollian solution space will have interesting and important ramifications to questions regarding black holes, which need to be thoroughly explored. Here we just mention two such ramifications. Our analysis clearly states that RT solution contributes neither to the boundary flux nor to the null boundary symplectic form. This implies that Hawking radiation in a pure gravity theory would be mainly dominated by the modes that are not of the RT type. Hence, {it} {naively} would not be relevant to the resolution of the information problem. {Additionally}, {the} appearance of {the} RT solution as the null direction in the solution space Carrollian geometry implies that we should label these geometries by charges {that} may be computed from our symplectic form. We hope to return to this point in an upcoming publication. 

{In addition to} our previous line of research, our analysis here was {partially} {motivated} by the recent paper  \cite{Ciambelli:2023mir}. In comparison to {our work}, that {paper} focuses on {the} off-shell symplectic form and {attempts}  to invert it to read off-shell Poisson brackets. {Similar to}  our analysis in section \ref{sec:corotating}, \cite{Ciambelli:2023mir}  also considers the non-rotating case ({which} more precisely should be called co-rotating). However, they do not include the ${\cal P}$ mode {in their analysis}. Therefore, the solution phase space they consider is not well-defined. {This can be readily observed from} our off-shell symplectic form \eqref{off-shell-symplectic}. As we see  ${\cal P}$ is conjugate to $\Omega \theta$. Therefore, to consistently truncate the system to the subspace without ${\cal P}$, one should also fix its conjugate, e.g. restricting to non-expanding case discussed in section \ref{sec:non-expanding}. We differ {from} \cite{Ciambelli:2023mir} in another {significant} way. We consider {the} on-shell symplectic form. Dealing with null boundaries and formulation of gravity on null-fronts, as illustrated in appendix \ref{appen-A}, it is physically meaningful (and necessary) to consider on-shell symplectic form and work with solution space. 

To state our main results, namely on-shell symplectic form \eqref{main-result} and the Carrollian nature of {the} null boundary solution space (cf. section \ref{sec:soln-space-carrollian}), we focused on the co-rotating sector in which the ``super-spin'' part ${\cal S}_A$ (and its off-shell canonical conjugate ${\cal U}^A$) have been turned off. We expect both of these results to extend to the most general solution space where ${\cal S}_A$ is also included. Explicitly, we expect an additional term in $\bO_{\text{\tiny bdy}}$ \eqref{main-result} which contains  $\slashed{\delta} {\cal S}_A\wedge \slashed{\delta} {\cal S}_B$ where $\slashed{\delta} {\cal S}_A$ is nonclosed 1-form and the nonclosed part is proportional to the divergence of the news, the (Bondi) angular momentum aspect. We also expect the bulk part of solution space {to} still remains Carrollian. 

In our analysis, we did not focus on the surface charges and {only} briefly mentioned them in appendix \ref{appen-charge}. These charges and their variations may be used to define horizon memory effect \cite{Adami:2021nnf} and null surface thermodynamics \cite{Adami:2021kvx}. Our symplectic form analysis here and the notion of symplectic flux shed some further light on both of these effects. In particular, it would be instructive to study if {there} is a horizon memory effect involving RT solutions and the null direction on the Carrollian solution space. Our analysis here on null boundary ${\cal N}$, with slight modifications, can be extended to asymptotic null infinity. It would be interesting to study the memory effect involving RT solutions in that setting. 

Finally, we would like to comment {on the} possible relation between the two Carrollian geometries we have in our setting: Carrollian geometry on the $D-1$ dimensional null boundary and the one on the bulk solution space. One may argue that the Carrollian structure {of} the solution space is a consequence of studying the solution space in {the} presence of a null boundary. A careful inspection of the details of the construction of the null boundary solution space (see \cite{Adami:2021nnf}) reveals that $\Omega$ being along the  kernel vector in the bulk solution space stems from the fact that in metric \eqref{metric'} $V=0$ at ${\cal N}$. 
This latter is of course equivalent {to} dealing with a null boundary. If this intuition is correct, this means that replacing the null boundary with a timelike boundary, e.g. replacing horizon with the stretched horizon, should also lift {the} null direction in the solution space. In other words, {the} Carrollian structure of the solution space is a consequence of dealing with a null boundary and in a timelike boundary case we do not expect to see a Carrollian solution space. Explicitly, our intuition is that  $V_0=V(r=0)$ in the solution space is expected to play the role of speed of light in usual Lorentzian geometry, such that in $V_0\to 0$ limit the solution space becomes Carrollian.  It would be instructive to establish this point.

\section*{Acknowledgment}

We would like to thank authors of \cite{Ciambelli:2023mir} for email exchanges and especially Luca Ciambelli for discussions which triggered this work. We would also like to thank Lars Andersson, Glenn Barnich, Luca Ciambelli, Daniel Grumiller and Shing-Tung Yau  for discussions. MMShJ would like to thank ULB, Brussels where this work was started. Work of MMSHJ is partially supported by INSF SarAmadan grant. The work of HA is supported
by Beijing Natural Science Foundation under Grant No. IS23018 and by the National Natural Science Foundation of China under Grant No. 12150410311.
The work of HY is supported in part
by Beijing Natural Science Foundation under Grant No. IS23013.

\appendix  

\section{Symplectic analysis of massless scalar theory on the light front}\label{appen-A}

Let us consider flat spacetime in the light-cone coordinate system with the line element 
\begin{equation}
    \d s^2 = - 2 \d u \d v + \d{}\mathbf{x} \cdot \d{}\mathbf{x}\, ,
\end{equation}
where $u$ and $v$ are null coordinates, and $\mathbf{x}^A$ are coordinates on a constant $(u,v)$ $(D-2)$-surface. The action for a system of massless scalar fields $\phi_i, i=1,2,\cdots, N$, can be written as:
\begin{equation}
   S=\int \d u  \,   L[\phi_i],  \qquad L[\phi_i]=\int_{u=cte.} \d v \d{}^{D-2}x\,{\cal L},\qquad {\cal L}=G^{ij} \left[ \partial_v \phi_i \, \partial_u \phi_j - \frac{1}{2} \left( \partial_{\mathbf{x}} \phi_i\cdot \partial_{\mathbf{x}} \phi_j\right)\right] \, ,
\end{equation}
where $L[\phi_i]$ is the Lagrangian, ${\cal L}$ is the Lagrangian density, and $G^{ij}$ is the metric over the field space. While for {physically} interesting cases $G^{ij}$ {is} either a function of $x$ and/or fields $\phi_i$, for simplicity and as an illustrative case, we take $G^{ij}$ to be a constant matrix. In our case, we assume the field space metric $G^{ij}$ to be invertible and $G^{ij} G_{jk}=\delta^i{}_k$.   
For the case of our interest, the off-shell or on-shell bulk symplectic forms, the metric on the field space is indeed field-dependent. The analysis of this appendix will be completed by those in appendix \ref{appen-inverting}.

We take $u$ to be the light-cone time from the bulk viewpoint. The conjugate light-cone momentum and the canonical Hamiltonian read
\begin{subequations}
    \begin{align}
         \pi^i &= \frac{\partial \mathcal{L}}{\partial (\partial_u \phi_i)} = G^{ij}\partial_v \phi_j\, ,\label{momentum-c}\\
          \qquad  \mathcal{H}_c &= \frac{1}{2} G^{ij} \partial_{\mathbf{x}} \phi_i\cdot \partial_{\mathbf{x}} \phi_j \, .
    \end{align}
\end{subequations}
As it has been discussed in \cite{Gonzalez:2023yrz,Alexandrov:2014rta}, we shall consider \eqref{momentum-c} a primary constraint, i.e.
\begin{equation}\label{momentum-on-shell-const}
    \chi^i := \pi^i - G^{ij}\partial_v \phi_j \approx 0\, .
\end{equation}
Hence we can rewrite the action as
\begin{equation}
   S[\phi_i,\pi^i;\lambda_i]=\int \d u \d v \d{}^{D-2}x\,\left[ \partial_u \phi_i \, \pi^i - \mathcal{H}_T \right]\, , \qquad \mathcal{H}_T = \mathcal{H}_c+ \lambda_i \, \chi^i\, ,
\end{equation}
where $\mathcal{H}_T$ is the total Hamiltonian density, which on a constant $u$ surface, say $u=0$, is given as
\begin{equation}
    H_{T}:= \int_{u=0} \d v \d{}^{D-2}x\ \mathcal{H}_{T} = \int_{u=0} \d v \d{}^{D-2}x \left[ \frac{1}{2} G^{ij} \partial_{\mathbf{x}} \phi_i\cdot \partial_{\mathbf{x}} \phi_j +\lambda_i \, \chi^i \right]\, .
\end{equation}
and $\lambda_i$ are Lagrangian multipliers. Hence the Hamilton equations read as
\begin{subequations}
    \begin{align}
        & \partial_u \phi_i =\frac{\delta H_T}{\delta \pi^i}= \lambda_i \, ,\\
        & \partial_u \pi^i =-\frac{\delta H_T}{\delta \phi_i}= G^{ij}\left(\partial_{\mathbf{x}}^2\phi_j - \partial_v \lambda_j\right)\, .
    \end{align}
\end{subequations}

Before proceeding further, we {want to emphasize}  the differences between the light-front analysis and the usual case where the Cauchy data (solution phase space data) are given on spatial constant time slices. In the light-front case, the ``Cauchy'' (boundary) data are given as $\phi^{(0)}_i(v,\mathbf{x})=\phi_i(u=0; v,\mathbf{x})$, and the momentum conjugate is $\pi_i=\partial_v\phi_i$. Therefore, the relation between the momenta and their conjugate fields at {the} null boundary $u=0$ {provides} a relation among the ``Cauchy'' data. This is {in contrast to}  the spatial boundary case, where momenta and their conjugate fields at the spatial boundary are independent variables on the solution phase space. { In the spatial boundary case,  \eqref{momentum-on-shell-const} {represent}   constraints on the solution phase space. These constraints, as we discuss below, are second-class constraints.

The symplectic potential can be inferred by taking the variation of the action. The symplectic potential on a constant $u$ slice at the null boundary,  {for example,} $u=0$, is {given by:}
\begin{equation}\label{symplectic-pot-scalar-1}
    \boldsymbol{\Theta} =\int_{u=0} \d{} v \d{}^{D-2} x \, \phi_i \, \delta \pi^i\, ,
\end{equation}
and hence the symplectic form can be {expressed} as:
\begin{equation}\label{off-shell-sym-form-scalar}
    \bO = \int_{u=0} \d{} v \d{}^{D-2} x \,  \delta \phi_i \wedge \delta \pi^i\, .
\end{equation}
Now we can immediately infer the Poisson bracket {as follows:}
\begin{equation}\label{PB-scalar}
\begin{split}
 \{\phi_i(v, \mathbf{x}),\phi_j(v', \mathbf{x}')\}_{\text{\tiny PB}}  &=0\, ,\\   
    \{\phi_i(v, \mathbf{x}),\pi^j(v', \mathbf{x}')\}_{\text{\tiny PB}}&=\delta_i{}^j\ \delta(v-v')\delta^{D-2}(\mathbf{x}-\mathbf{x}')\, ,\\
    \{\pi^i(v, \mathbf{x}),\pi^j(v', \mathbf{x}')\}_{\text{\tiny PB}}&=0\, .
\end{split}\end{equation}

Defining the smeared constraints {as} $X= \int \d v \d{}^{D-2}x\ f_i \chi^i$, where $f_i(v,\mathbf{x})$ are test functions that vanish at the boundary, one finds the Poisson bracket for $\chi^i$ {as}:
\begin{equation}
    \{\chi^i(v, \mathbf{x}),\chi^j(v', \mathbf{x}')\}_{\text{\tiny PB}} =-2 G^{ij}\ \partial_{v} \delta(v-v')\delta^{D-2}(\mathbf{x}-\mathbf{x}')\, .
\end{equation}
The consistency condition leads to nothing new
\begin{equation}
        \partial_u \chi^i  = \{ \chi^i, H_{T}\}_{\text{\tiny PB}} \approx 0\, .
\end{equation}
\paragraph{Dirac bracket.} To compute the Dirac bracket, we need to inverse the Poisson brackets of the second-class constraints, namely $\Delta^{ij}(v,\mathbf{x};v',\mathbf{x}')=\{\chi^i(v, \mathbf{x}),\chi^j(v', \mathbf{x}')\}_{\text{\tiny PB}}$,
\begin{equation}
    (\Delta^{-1})_{ij}(v,\mathbf{x};v',\mathbf{x}')= {\frac14} G_{ij}\ H(v-v')\ \delta^{D-2}(\mathbf{x}-\mathbf{x}')\, ,
\end{equation}
where 
\begin{equation}\label{step-function}
    H(v-v'):=\left\{\begin{array}{cc} \frac{(v-v')}{|v-v'|} & \qquad v\neq v'\\ 0 & \qquad v=v' \end{array}\right.
\end{equation}
is the Heaviside step function. Note that with the above definition $\partial_v H(v-v')=2\delta(v-v')$.  

From this, one can {determine} the Dirac bracket of two arbitrary functions in the phase space, $f$ and $g$, as follows:
\begin{equation}
    \begin{split}
        &\{f(v,\mathbf{x}),g(v',\mathbf{x}')\}_{\text{\tiny{DB}}}=\{f(v,\mathbf{x}),g(v',\mathbf{x}')\}_{\text{\tiny{PB}}}\\
        &-\int\d{}\tilde{v}\d{}^{D-2}\tilde{x}\int\d{}\bar{v}\d{}^{D-2}\bar{x}\{f(v,\mathbf{x}),\chi^i(\tilde{v},\tilde{\mathbf{x}})\}_{\text{\tiny{PB}}}(\Delta^{-1})_{ij}(\tilde{v},\tilde{\mathbf{x}};\bar{v},\bar{\mathbf{x}})\{\chi^j(\bar{v},\bar{\mathbf{x}}),g(v',\mathbf{x}')\}_{\text{\tiny{PB}}}\, .
    \end{split}
\end{equation}
The Dirac bracket of two scalar fields is then given as
\begin{equation}\label{basic-bracket-scalar}
\boxed{    \{\phi_i(v,\mathbf{x}),\phi_j(v',\mathbf{x}')\}_{\text{\tiny{DB}}}= {\frac14}G_{ij}\ H(v-v')\ \delta^{D-2}(\mathbf{x}-\mathbf{x}')\, .}
\end{equation}

\paragraph{On-shell symplectic form and Poisson brackets on {the} reduced phase space.} One can {directly} obtain the bracket \eqref{basic-bracket-scalar} from equation \eqref{symplectic-pot-scalar-1} without using {the} Dirac bracket procedure. To do so, we note that {by} imposing the $\chi^i\approx 0$ constraints, which are  {essentially} half of Hamilton's equations, and replacing the momenta {with} the derivatives of the fields, {we arrive at the} ``on-shell'' symplectic form:
\begin{equation}\label{on-shell-sym-form-scalar}
\begin{split}
    \bO_{\text{\tiny{on-shell}}} &= \int \d{} v \d{}^{D-2} x \, G^{ij}  \delta \phi_i \wedge \partial_v \delta\phi_j\, \\
&=\frac{1}{2}\int \d{} v \d{}^{D-2} x \int \d{} v' \d{}^{D-2} x' \,   \delta \phi_i(v,\mathbf{x}) \bO^{ij} (v,\mathbf{x};v',\mathbf{x}')\wedge \delta\phi_j(v',\mathbf{x}')\, ,
\end{split}\end{equation}
where 
\begin{equation}\label{symp-form-scalar}
    \begin{split}
    \bO^{ij}(v,\mathbf{x};v',\mathbf{x}') =  - \bO^{ji}(v',\mathbf{x}';v,\mathbf{x})= 2 G^{ij}\ \partial_{v} \delta(v-v') \delta^{D-2} (\mathbf{x}-\mathbf{x}')\, .
    \end{split}
\end{equation}
In the {terminology of typical constrained systems}, $\bO_{\text{\tiny{on-shell}}}$ {represents}  the symplectic form over the reduced phase space obtained after solving and imposing constraints. One can invert the above symplectic potential over this reduced phase space to find:
\begin{equation}\label{symp-form-inv-scalar}
     (\bO^{-1})_{ij} (v,\mathbf{x};v',\mathbf{x}')= {\frac14} G_{ij} \ H(v-v')\ \delta^{D-2}(\mathbf{x}-\mathbf{x}')\, .
\end{equation}
It is {evident }  that the  symplectic form  above directly yields the bracket \eqref{basic-bracket-scalar}. 

\section{A quick review of Carrollian geometry}\label{appen-carrollian}
Let us consider a Lorentizan $d$-dimensional manifold equipped {with}  a degenerate metric $g_{ab}$ of rank $(d-1)$. {We can use coordinate}  $x^a= \{v,x^A\}, \ A=1,2,\cdots, d-1$ on the given manifold. The generic line element in this case takes the form:
\begin{equation}\label{Carroll-metric}
    \d s^2 =g_{ab} \d x^a \d x^b = g_{AB} (\d x^A+ U^A \d v)(\d x^B+ U^B \d v)\, .
\end{equation}
To describe Carrollian geometry, we need to define the kernel of the metric
\begin{equation}\label{Kernel-Carroll}
   K^a\partial_a = \alpha \left( \partial_v - U^A \partial_A \right)\, ,  \qquad  \qquad  K^a g_{ab}= 0 \, ,
\end{equation}
which is the Carrollian kernel vector and the Ehresmann connection 
\begin{equation}
    E_a \d x^a = \frac{1}{\alpha}\, \d v+ E_A (\d x^A+ U^A \d v) \, , \qquad \qquad K^a E_a=1 \, .
\end{equation}
The metric \eqref{Carroll-metric} and \eqref{Kernel-Carroll} define a Carrollian geometry, which is naturally  supplemented with the Ehresmann connection. 

The minors of $g_{ab}$ is proportional to the product $K^a K^b$ \cite{Henneaux:1979vn,Henneaux:2021yzg}, i.e. 
\begin{equation}
    \mathfrak{g}^{ab}=\beta^2 \, K^a K^b\, .
\end{equation}
The determinant of $g_{ab}$ vanishes, however, its determinant is replaced by the density $\beta$. While the metric is degenerate and hence non-invertible, we can still raise indices by exploiting the Ehresmann connection. We may define a symmetric $(2,0)$-type tensor $h^{ab}$ such that:
\begin{equation}
    h^{ac} g_{cb}= \delta^a_b - K^a E_b\, .
\end{equation}
To fully determine $h^{ab}$, we would need to impose an additional condition $h^{ab} E_a E_b=0$.

Let us set $\alpha=1, \, E_A=0$, for simplicity. Then the components of ${h}^{ab}$ read
\begin{equation}
    h^{vv}=0 \, , \qquad h^{vA}= 0\, , \qquad h^{AB}= g^{AB} \, .
\end{equation}
This justifies {the} discussion on the invertable part of {the} bulk symplectic form in item \ref{Item-V} in section \ref{sec:corotating}.

\section{Null surface boundary symmetry generators}\label{appen-sym-gen}
The following vector field {represents} the null surface symmetries {that} preserves the form of {the} metric and moves us into the solution space
\begin{equation}\label{sym-generator}
    \xi =T\, \partial_v + r \left(\mathcal{D}_v T -  W  \right) \partial_r + \left( Y^A - {\cal U}^AT -r \eta  \partial^A T\right)\partial_A + \mathcal{O}(r^2)\, ,
\end{equation}
in which $T(v,x^A)$,  $W(v,x^A)$, and $Y^{A}(v,x^B)$ are codimension 1 symmetry generators of the causal boundary.
This vector field keeps $r=0$ a null surface and generates the following variations over the solution phase space functions
\begin{subequations}\label{delta-charges'}
    \begin{align}      
        \delta_\xi \eta &=  T \mathcal{D}_{v} \eta +2 \eta {\cal D}_v T-W\eta  +Y^A \partial_A \eta\,,\\ 
        \delta_\xi\gamma_{AB} &=T {\cal D}_{v}\gamma_{AB}+2\nabla_{\langle A} Y_{B \rangle}\\
      \delta_{\xi}{\cal J}_{A}&= T {\cal D}_v {\cal J}_A+\mathcal{L}_{Y}{\cal J}_{A}+\Omega \big[\partial_{A}W- {\Gamma}\partial_{A}T-{2}{N_{A B}}\partial^{B}T\big]
    \,,\label{delta-JA}\\
               \delta_{\xi}\Gamma &=-{\cal D}_{v}(W-\Gamma T)+{Y}^{A}\partial_{A}\Gamma\,,\label{delta-Gamma} \\ 
               \delta_{\xi}\mathcal{N}_{AB} &={\cal D}_v (T{\cal N}_{AB})+{\cal L}_{ Y}{\cal N}_{AB}-\frac{2\nabla_{C}{Y}^C}{D-2}\mathcal{N}_{AB}
    \, ,\label{delta-NAB}
\end{align}
\end{subequations}
and ${\cal L}_{Y}$ denotes the Lie derivative along ${Y}^A$.

\paragraph{Transformation of charge densities and their canonical conjugates.}

From the above, one may compute variations of $\varpi, \Omega ;  \vartheta, {\cal P}; {\cal U}^A, {\cal S}_A$,
\begin{subequations}\label{delta-charges-gamma}
    \begin{align}   
    \delta_{\xi} \varpi &=\mathcal{D}_{v}(T \varpi)+Y^{A}\partial_{A}\varpi-\partial_{v}W+\mathcal{U}^{A}\partial_{A}T \partial_{v}\mathcal{P}+\partial_{A}\mathcal{P}\partial_{v}(Y^{A}-\mathcal{U}^{A}T)\, , \\
    \delta_{\xi}\Omega &= T{\cal D}_v \Omega+ \partial_A(\Omega Y^A)\, ,\\ 
     \delta_{\xi} \vartheta &=\partial_{v}[\vartheta T-T\partial_{A}(\Omega \mathcal{U}^{A})]+\nabla_{A}(\vartheta Y^{A}+\Omega \partial_{v}Y^{A})\,  =\partial_v (\delta_{\xi}\Omega)\, ,\\
    \delta_\xi {\cal P} & \approx T\varpi +(Y^{A}-T\mathcal{U}^{A})\partial_{A}{\cal P}-W +\frac{2T}{\theta} {\cal N}^2,\\
     \delta_\xi \mathcal{U}^{A} &={\cal D}_v {Y}^A \,,\label{delta-UA} \\
    \delta_{\xi}{\cal S}_{A}&\approx \nabla_{B}({Y}^{B}\mathcal{S}_{A})+\mathcal{S}_{B}\nabla_{A}{Y}^{B}+2\Omega \nabla_{A}(T\theta^{-1}\mathcal{N}^2)-2\nabla_A^{B}(T \Omega \mathcal{N}_{AB})\, ,
\end{align}
\end{subequations}
where $\gamma_{AB}, {\cal P}, {\cal N}_{AB}$ are defined in \eqref{gamma-P-def}, \eqref{S-varpi-vartheta} and 
\begin{equation}
    {\cal D}_v {\cal N}_{AB}:=\partial_{v} \mathcal{N}_{AB}-\mathcal{L}_{\mathcal{U}}\mathcal{N}_{AB} + \frac{2}{D-2}\, \mathcal{N}_{AB}\, \nabla_{C}\mathcal{U}^{C}\, .
\end{equation}
Note that the variation of the superspin does not depend on $W$ (compare {it} with that of \eqref{delta-JA}). $\delta_\xi {\cal P}$ and $\delta_\xi {\cal S}_A$ are written ``on-shell'' using {the} Raychaudhuri \eqref{Raych} and Damour \eqref{Damour} equations. For completeness, we also present variations  $\slashed{\delta}_\xi \hat{\cal P}, \slashed{\delta}_\xi \hat{\gamma}_{AB}$ defined in \eqref{delta-slashes}
\begin{subequations}
    \begin{align}
         & \slashed{\delta}_\xi \hat{\cal P}\approx -W + \varpi T +(Y^{A}-T\mathcal{U}^{A})\partial_{A}\mathcal{P}-2\theta^{-1}\mathcal{N}^{A}_{B}\nabla_{A}Y^{B}\, ,\\
         & \slashed{\delta}_\xi \hat{\gamma}_{AB} =  2\nabla_{\langle A}Y_{B \rangle}-\frac{2\mathcal{N}_{AB}}{\Omega\theta}\nabla_{C}(\Omega Y^{C})\, .\label{delta-xi-hat-gamma}
    \end{align}
\end{subequations}
Interestingly, note that $\slashed{\delta}_\xi \hat{\gamma}_{AB}$ has no $T,W$ dependence. The first term in \eqref{delta-xi-hat-gamma} is variation of a $D-2$ dimensional symmetric 2 tensor and the term proportional to ${\cal N}^{AB}$ is stemming from the non-closedness of $\slashed{\delta}\hat{\gamma}$.

\section{Charge analysis}\label{appen-charge}
{Given the symplectic form, one can define the surface charge variation as $\slashed{\delta}Q_{\xi}= \bO[\delta_\xi g , \delta g ; g]$ which becomes a surface integral over co-dimension 2 surface on-shell \cite{Lee:1990nz,Iyer:1994ys}.}
Surface charge variation associated with symmetry generator \eqref{sym-generator} is \cite{Adami:2021nnf, Adami:2021kvx}
\begin{equation}\label{surface-charge-01}
        \slashed{\delta} Q_{\xi}= \frac{1}{16\pi G} \int_{{\cal N}_b} \left[ \left(W-\Gamma T \right)\delta\Omega +Y^{A}\delta{\cal J}_{A}  
        +  T\, \Omega \, \theta \, \slashed{\delta}\hat{{\cal P}}   \right] \, ,
\end{equation}
or in the Heisenberg-direct sum slicing, the charge variation can be split into the integrable part
\begin{equation}
     \tilde{Q}^{\text{I}}_{\xi}= \frac{1}{16\pi G} \int_{{\cal N}_b}  \left( \tilde{W} \Omega+Y^{A}  \mathcal{S}_{A}+\tilde{T} \mathcal{P}\right)\, ,
\end{equation}
and the flux
\begin{equation}\label{100}
    \mathbf{F}_\xi (\delta g)= -\frac{1}{16\pi G} \int_{{\cal N}_b}   \left[ \tilde{T}-\partial_{C}(\Omega Y ^{C})\right]\theta^{-1} \mathcal{N}^{AB}\delta\gamma_{AB}\, ,
\end{equation}
where \cite{Adami:2021nnf}
\begin{equation}\label{hat-slicing-10}
    \tilde{W}=W-\Gamma T-Y^A \partial_A\mathcal{P}, \qquad \tilde{T}=\Omega\theta  T+\partial_A(\Omega Y^A)\, .
\end{equation}
We note that $\delta_\xi\Omega= \tilde{T}$ and hence when $Y^A=0$, $\mathbf{F}_\xi (\delta g)=     \mathbf{F}_{\text{\tiny bdy}}\big|_{\delta\Omega=\delta_\xi\Omega}$, where $    \mathbf{F}_{\text{\tiny bdy}}$ is the boundary symplectic flux defined in \eqref{F-bdry}. This clarifies the relation between the charge flux and the symplectic flux.\footnote{Note that the inner product of the vector field induced by $\xi$ and the symplectic flux \eqref{F-bdry} is {as follows:}
$$  I_{\xi} \mathbf{F}_{\text{\tiny bdy}} = -\frac{1}{16\pi G} \oint_{\mathcal{N}_b}   \frac{{\cal N}^{AB}}{\theta} (\delta_\xi \Omega \delta\gamma_{AB}-\delta \Omega \delta_\xi \gamma_{AB})=-\frac{1}{16\pi G} \int_{{\cal N}_b}   \left[ \tilde{T}\frac{{\cal N}^{AB}}{\theta}\slashed{\delta}\hat{\gamma}_{AB}-2\delta\Omega \left(\frac{{\cal N}^{AB}}{\theta}\nabla_A Y_B-\frac{{\cal N}^2}{\theta^2} \nabla_C(\Omega Y^C)\right)\right]\, .
$$
}

Now, by using the adjusted bracket proposed by Barnich and Troessaert \cite{Barnich:2011mi}, the charge algebra can be read as follows
\begin{subequations}\label{Heisenberg-direct-sum-algebra}
    \begin{align}
        &\{\Omega(v,\mathbf{x}),\Omega(v,\mathbf{x}')\}=\{\mathcal{P}(v,\mathbf{x}),\mathcal{P}(v,\mathbf{x}')\}=0\, ,\\
        &\{\Omega(v,\mathbf{x}),\mathcal{P}(v,\mathbf{x}')\}=16\pi G\delta^{D-2}\left(\mathbf{x}-\mathbf{x}'\right)\, ,\\
        &\{\mathcal{S}_A(v,\mathbf{x}),\Omega(v,\mathbf{x}')\}=\{\mathcal{S}_A(v,\mathbf{x}),\mathcal{P}(v,\mathbf{x}')\}=0\, ,\\
        &\{\mathcal{S}_A(v,\mathbf{x}),\mathcal{S}_B(v,\mathbf{x}')\}=16\pi G\left(\mathcal{S}_{A}(v,\mathbf{x}')\partial_{B}-\mathcal{S}_{B}(v,\mathbf{x})\partial'_{A}\right)\delta^{D-2}\left(\mathbf{x}-\mathbf{x}'\right)\, .
    \end{align}
\end{subequations}
Note that the above are equal-$v$ charge brackets.  This $v$ may be taken to be the arbitrary value $v_b$. {Thus,} the above is the same as \eqref{bdry-brackets-1}, \eqref{bdry-brackets-2}.


\section{Details of the on-shell symplectic form computations }\label{appen-on-shell-symp-details}

In this appendix, we provide a detailed  of derivation of \eqref{main-result}, starting from \eqref{Omega-on-shell}. To this end, we note that \eqref{Omega-on-shell} may be written as: 
\begin{equation}\label{On-shell-Symp-1}
    \begin{split}
        16\pi G \ \bO_{\text{\tiny on-shell}} &=  \oint_{{\mathcal{N}_b}} \bigl(\delta \Omega \wedge \delta \mathcal{P}   \bigr) 
        + \frac12\int_{\mathcal{N}}\, \Omega\mathcal{G}^{ABCD} \biggl[\frac{4}{\Omega^2\theta^2}\, \mathcal{N}_{AB}\mathcal{N}_{CD} \delta\Omega \wedge \partial_{v}\delta \Omega+  \delta\gamma_{AB}\wedge \partial_{v}\delta \gamma_{CD}\\
         &-\frac{4}{\Omega\theta}\, \mathcal{N}_{AB}\, \delta\Omega\wedge\partial_{v}\delta\gamma_{CD} +\frac{1}{(D-3)\Omega}\, \gamma_{AB}\,\delta\Omega\wedge\partial_{v}\delta\gamma_{CD}
        +\frac{{8}}{\Omega\theta}\,  \mathcal{N}_{AE}\mathcal{N}^{E}_{B}\, \delta\Omega \wedge \delta \gamma_{CD}\biggr]\, .
    \end{split}
\end{equation}

Now, we  concentrate  on the bulk term, which is the integral over ${\cal N}$ in \eqref{On-shell-Symp-1}, and establish its relationship with $\bO_{\text{\tiny Bulk}}$ from \eqref{main-result}. Let us start from \eqref{Omega-bdry-bulk},
\begin{equation}
          {16\pi G\,}\bO_{\text{\tiny Bulk}} = \frac12\int_{\mathcal{N}}\, \Omega\mathcal{G}^{ABCD} \slashed{\delta}{\hat{\gamma}}_{AB}\wedge \partial_v \slashed{\delta}{\hat{\gamma}}_{CD}= \bO_1 +\bO_2+\bO_3+\bO_4\, , 
\end{equation}
where using \eqref{slashed-delta-gamma} we have
\begin{equation}\label{Omega-Bulk-derivation-1}
    \begin{split}
        \bO_1 &=\frac12\int_{\mathcal{N}}\, \Omega\mathcal{G}^{ABCD} \, \delta \gamma_{AB} \wedge \partial_v \delta \gamma_{CD}\, , \\
        \bO_4 &=\frac42\int_{\mathcal{N}}\, \Omega\mathcal{G}^{ABCD} \,\left(  \frac{\mathcal{N}_{AB}}{\Omega \, \theta} \, \delta \Omega\right)\wedge \partial_v \left(  \frac{\mathcal{N}_{CD}}{\Omega \, \theta} \,\delta \Omega\right)=\int_{\mathcal{N}} \frac{2\mathcal{N}^2}{\Omega \, \theta^2} \delta\Omega\wedge\partial_v {\delta}\Omega\, , \\
        \bO_2 &=-\int_{\mathcal{N}}\, \Omega\mathcal{G}^{ABCD} \, \delta \gamma_{AB}\wedge \partial_v \left(  \frac{\mathcal{N}_{CD}}{\Omega \, \theta} \,\delta \Omega\right)\, ,\\
        \bO_3 &=-\int_{\mathcal{N}}\, \Omega\mathcal{G}^{ABCD} \,\left( \frac{\mathcal{N}_{AB}}{\Omega \, \theta} \, \delta \Omega\right)\wedge \partial_v \delta \gamma_{CD}\, .
    \end{split}
\end{equation}
Straightforward manipulations yield
\begin{equation}
    \begin{split}
        \bO_2&=-\oint_{{\mathcal{N}_b}} \Omega\mathcal{G}^{ABCD} \, \delta \gamma_{AB}\wedge   \frac{\mathcal{N}_{CD}}{\Omega \, \theta} \,\delta \Omega+ \int_{\mathcal{N}}\, \partial_v\left(\Omega\mathcal{G}^{ABCD} \, \delta \gamma_{AB}\right)\wedge  \left(  \frac{\mathcal{N}_{CD}}{\Omega \, \theta} \,\delta \Omega\right)\\
        &=-\oint_{{\mathcal{N}_b}} \frac{\mathcal{N}^{AB}}{\theta} \delta \gamma_{AB}\wedge\delta \Omega+ \int_{\mathcal{N}}\,\Biggl\{ {\cal N}^{AB}\delta\gamma_{AB}\wedge\delta\Omega+\frac{\mathcal{N}_{CD}}{\theta} \,\left[\partial_v(\mathcal{G}^{ABCD}) \, \delta \gamma_{AB}+ \delta \partial_v\gamma_{CD}\right]\wedge \delta \Omega \Biggr\},\\
        \bO_3&=-\int_{\mathcal{N}}\, \Omega\mathcal{G}^{ABCD} \,\left( \frac{\mathcal{N}_{AB}}{\Omega \, \theta} \, \delta \Omega\right)\wedge \partial_v \delta \gamma_{CD} = \int_{\mathcal{N}}\frac{\mathcal{N}^{AB}}{\theta}\delta \partial_v\gamma_{AB}\wedge \delta \Omega\, .
    \end{split}
\end{equation}
Recalling the definition of the Wheeler-DeWitt metric \eqref{WD-metric} yields
\begin{equation}
    \begin{split}
        \int_{\mathcal{N}}\,\frac{\mathcal{N}_{CD}}{\theta} \,\partial_v(\mathcal{G}^{ABCD}) \, \delta \gamma_{AB}\wedge \delta \Omega=-\int_{\mathcal{N}}\ \frac{4}{\theta}\mathcal{N}^{AC}\mathcal{N}_{C}{}^B \delta \gamma_{AB}\wedge \delta \Omega\,.
    \end{split}
\end{equation}
Therefore, we get
\begin{equation}
    \begin{split}
        \bO_2+\bO_3=&-\oint_{\mathcal{N}_{b}} \frac{\mathcal{N}^{AB}}{\theta} \delta \gamma_{AB}\wedge\delta \Omega\\
        &+ \int_{\mathcal{N}}\,\Biggl\{ {\cal N}^{AB}\delta\gamma_{AB}\wedge\delta\Omega-\frac{4}{\theta}\mathcal{N}^{AC}\mathcal{N}_{C}{}^B \delta \gamma_{AB}\wedge \delta \Omega+ 2\frac{\mathcal{N}^{AB}}{\theta}\delta \partial_v\gamma_{AB}\wedge \delta \Omega\biggr\}\, .
    \end{split}
\end{equation}
Using definition of ${\cal N}_{AB}$, we also note that
\begin{equation}
\int_{\mathcal{N}} \gamma^{AB}\,\delta\Omega\wedge\partial_{v}\delta\gamma_{AB}= {-}\int_{\mathcal{N}} \partial_v\gamma^{AB}\,\delta\Omega\wedge\delta\gamma_{AB}=+{2}\int_{\mathcal{N}} {\cal N}^{AB}\,\delta\Omega\wedge\delta\gamma_{AB}\,.
\end{equation}
Putting all the above together and summing $\bO_1+\bO_2+\bO_3+\bO_4$ yields the desired result. 

\section{Derivation of bulk off-shell and on-shell Poisson brackets}\label{appen-inverting}
In this appendix, we first {provide the}  details of {the} derivation of \eqref{PB-NN''}, and {then we will demonstrate}  how to compute on-shell Poisson brackets of the bulk modes in the sector where the bulk symplectic form is invertible. 

\subsection{Off-shell bulk Poisson brackets, \texorpdfstring{\eqref{PB-NN''}}{} }\label{appen-eq(3-5)}
The symplectic form leads to off-shell Poisson brackets
\begin{subequations}\label{off-shell-Poisson-brackets'}
    \begin{align}
        & \left\{ \mathcal{U}^A(x_1) , \mathcal{S}_B(x_2)\right\}_{\text{\tiny PB}}= 16 \pi G \, \delta^A_B \, \delta^{D-1}(x_1-x_2)\, , \label{PB-a'}\\
        &  \left\{ \Omega(x_1) , \varpi(x_2)\right\}_{\text{\tiny PB}}= 16 \pi G  \, \delta^{D-1}(x_1-x_2)\, , \label{PB-b'}\\
        &  \left\{ \vartheta(x_1) , \mathcal{P}(x_2)\right\}_{\text{\tiny PB}}= 16 \pi G  \, \delta^{D-1}(x_1-x_2)\, , \label{PB-c'}\\
        & \left\{ \mathcal{N}_{AB}(x_1) , \Omega(x_2) \, \gamma^{CD}(x_2)\right\}_{\text{\tiny PB}}= 16 \pi G \, \delta^C_{(A} \delta^D_{B)} \, \delta^{D-1}(x_1-x_2)\, . \label{PB-N'}
    \end{align}
\end{subequations}
Poisson brackets among {the following} second-class constraints,
\begin{equation}\label{}
    C_1:=\det{\gamma}-1\approx 0,\qquad C_2:=\gamma^{AB}{\cal N}_{AB}\approx 0,
\end{equation}
read as
\begin{subequations}
    \begin{align}
        & \left\{ C_1(x_1), C_1(x_2)\right\}=0 \, ,\\
        & \left\{ C_1(x_1), C_2(x_2)\right\}=\frac{16 \pi G \, (D-2) }{\Omega(x_1)}\, \det{\gamma}(x_1) \,  \delta^{D-1}(x_1-x_2)\, ,\\
        & \left\{ C_2(x_1), C_2(x_2)\right\}= 0\, ,
    \end{align}
\end{subequations}
where the following equation was used
\begin{equation}
    \frac{\delta \det{\gamma}}{\delta \gamma^{AB}}= -\det{\gamma} \, \gamma_{AB}
\end{equation}
The above can be written as a matrix
\begin{equation}
    \Delta_{ij}(x_1,x_2) \approx \frac{16 \pi G \, (D-2) }{\Omega(x_1)}\,   \delta^{D-1}(x_1-x_2)
    \begin{pmatrix}
      0  & 1\\
      -1  & 0
    \end{pmatrix}
\end{equation}
where $i,j=1,2$. We may use the equation
\begin{equation}
    \int_{\mathcal{N}}''\Delta_{ik}(x,x'')(\Delta^{-1})^{kj}(x'',x')= \delta^j_{i} \,  \delta^{D-1}(x-x')
\end{equation}
to define the inverse of $\Delta_{ij}(x_1,x_2)$ and hence
\begin{equation}
    (\Delta^{-1})^{ij}(x_1,x_2) = \frac{\Omega(x_1)}{16 \pi G \, (D-2)} \, \delta^{D-1}(x_1-x_2)\, 
    \begin{pmatrix}
      0  & -1\\
      1  & 0
    \end{pmatrix}
\end{equation}

\begin{equation}
    \begin{split}
        \left\{ \mathcal{N}_{AB}(x) , \gamma^{CD}(x')\right\}_{\text{\tiny{DB}}}=&\left\{ \mathcal{N}_{AB}(x) , \gamma^{CD}(x')\right\}_{\text{\tiny{PB}}}\\
        &-\int_\mathcal{N}''\int_\mathcal{N}'''\{\mathcal{N}_{AB}(x),C_i(x'')\}_{\text{\tiny{PB}}}(\Delta^{-1})^{ij}(x'';x''')\{C_j(x'''),\gamma^{CD}(x')\}_{\text{\tiny{PB}}}\, \\
        =&\ \frac{16 \pi G}{\Omega(x)} \, \delta^C_{(A} \delta^D_{B)} \, \delta^{D-1}(x-x')\\
        &+\int_\mathcal{N}''\frac{\Omega(x'')}{16 \pi G \, (D-2)} \,\{\mathcal{N}_{AB}(x),C_1(x'')\}_{\text{\tiny{PB}}} \{C_2(x''),\gamma^{CD}(x')\}_{\text{\tiny{PB}}}\, \\
        =&\ \frac{16 \pi G}{\Omega(x)} \left[ \, \delta^C_{(A} \delta^D_{B)}-\frac{1}{ (D-2)} \, \gamma_{AB}(x) \gamma^{CD}(x) \right]  \delta^{D-1}(x-x')
    \end{split}
\end{equation}
then
\begin{equation}\label{PB-NN'}
    \left\{ \mathcal{N}_{AB}(x) ,  \gamma^{CD}(x')\right\}_{\text{\tiny DB}}= \frac{16\pi G}{\Omega(x) }
    \,\left(\delta^C_{(A} \delta^D_{B)} -\frac{1}{D-2}\gamma_{AB}\gamma^{CD}\right) \delta^{D-1}(x-x')\, .
\end{equation}
Using the identity
\begin{equation}
    \begin{split}
        \left\{ \mathcal{N}_{AB}(x) ,  \gamma^{CD}(x')\right\}_{\text{\tiny DB}} = &\ \left\{ \mathcal{N}_{AB}(x) ,  \gamma^{CE}(x')\gamma^{DF}(x')\gamma_{EF}(x')\right\}_{\text{\tiny DB}}\\
        =&\ \left\{ \mathcal{N}_{AB}(x) ,  \gamma_{EF}(x')\right\}_{\text{\tiny DB}}  \gamma^{CE}(x')\gamma^{DF}(x') + \left\{ \mathcal{N}_{AB}(x) ,  \gamma^{CE}(x')\right\}_{\text{\tiny DB}}  \gamma^{DF}(x')\gamma_{EF}(x') \\
        & + \left\{ \mathcal{N}_{AB}(x) ,  \gamma^{DF}(x')\right\}_{\text{\tiny DB}}  \gamma^{CE}(x')\gamma_{EF}(x')
    \end{split}
\end{equation}
one can readily check \eqref{PB-NN''}.

\subsection{On-shell bulk Poisson brackets}\label{appen-bulk-on-shell}

In the example of scalar fields discussed in Appendix \ref{appen-A}, the metric on the field space $G^{ij}$ was considered to be field-independent. In our case, however, the invertible part of the bulk symplectic form 
is spanned by $\gamma_{AB}$, {with}  $\Omega(x)$ {being a fixed function.}  Therefore
\begin{equation}\label{Bulk-I-1}
    \begin{split}
         \bO^{\text{\tiny{I}}}_{\text{\tiny Bulk}} & = \frac{1}{32 \pi G}\int_{\mathcal{N}} \delta^{A}_{(C}\delta^B_{D)} \Omega\,   \partial_v {\delta}{{\gamma}}_{AB} \wedge {\delta}{\gamma}^{CD}\\
          & :=\int_{\mathcal{N}}\int_{\mathcal{N}}' \bO^{ABCD}(x,x')   {\delta}{{\gamma}}_{AB}(x) \wedge {\delta}{\gamma}_{CD}(x') \, ,
        \end{split}
\end{equation}
where using $\delta\gamma^{CD}=-\gamma^{CE}\gamma^{DF}\delta\gamma_{EF}$, we obtain
\begin{equation}\label{bO-ABCD-xx'}
     \bO^{ABCD}[x;x'] =\frac{\sqrt{\Omega(x)\Omega(x')}}{64 \pi G} {\cal G} ^{ABCD}[x;x']{\left(\partial_{v}\delta(v-v')-\partial_{v'}\delta(v'-v)\right)}\delta^{D-2}(\mathbf{x}-\mathbf{x}')\, ,
\end{equation}
where ${\cal G} ^{ABCD}[x;x']$ is the \textit{point-split WdW metric} \eqref{point-split-WdW} which satisfies the following equations
\begin{subequations}
    \begin{align}
        & \gamma_{AB}(x){\cal G} ^{ABCD}[x;x']= {\cal G} ^{ABCD}[x;x']\gamma_{CD}(x')=0 \, , \\
        & {\cal G} ^{ABCD}[x;x']={\cal G} ^{CDAB}[x';x]={\cal G} ^{BACD}[x;x']={\cal G} ^{ABDC}[x;x']\, .
    \end{align}
\end{subequations}
Let us denote the inverted antisymmetric bivector by $\bohm_{ABCD}$, 
\begin{equation}\label{Bulk-bivector}
    \begin{split}
         \bohm^{\text{\tiny{I}}}_{\text{\tiny Bulk}}  :=\int_{\mathcal{N}}\int_{\mathcal{N}}' \bohm_{ABCD}[x;x']  \frac{\delta}{{\delta}{{\gamma}}_{AB}(x)}\vee \frac{\delta}{{\delta}{\gamma}_{CD}(x')} \, ,
        \end{split}
\end{equation}
such that $\bohm_{ABCD}[x;x']$ is trace-free, that is
\begin{equation}
    \gamma^{AB}(x)\bohm_{ABCD}[x;x']= \bohm_{ABCD}[x;x']\gamma^{CD}(x')=0\, ,
\end{equation}
and it is defined via\footnote{One can verify that 
$
     \int^{''}_{\cal N} \bohm_{CDEF}[x;x'']\bO^{EFAB}[x'';x']= 
    P^{AB}_{CD}\ \delta^{D-1}(x-x')
$ 
is equivalent to \eqref{Omega-inverse'}.}
\begin{equation}\label{Omega-inverse'}
     \int^{''}_{\cal N} \ \bO^{ABEF}[x;x'']\bohm_{EFCD}[x'';x']= 
    P^{AB}_{CD}\ \delta^{D-1}(x-x')\, .
\end{equation}
Then by definition, the on-shell Poisson bracket is 
\begin{equation}\label{gamma-PB-appen}
\boxed{    \left\{ \gamma_{AB}(x_{_1}) ,  \gamma_{CD}(x_{_2})\right\}= \bohm_{ABCD}[x_{_1};x_{_2}]\, }
\end{equation}
Similarly to the analysis of Appendix \ref{appen-eq(3-5)}, {when} working with trace-free $\bohm_{ABCD}[x;x']$, {it is important to note that}  the constraints  \eqref{C1-C2} are already implemented. {This is} because $C_2 \propto \partial_v C_1$ and hence $C_2$ is already satisfied ``on-shell'' if $C_1$ holds and $C_1$ is guaranteed by the traceless property of $\bohm_{ABCD}[x;x']$. 
To solve for $\bohm_{ABCD}(x_{_1};x_{_2})$, recalling analysis in appendix \ref{appen-A} and inspired by the point-split WdW metric \eqref{point-split-WdW}, one can immediately observe the following form
\begin{equation}\label{Bohm-appen-general'}
   \boxed{ \bohm_{ABCD}[x;x']= \frac{16\pi G}{\sqrt{\Omega(x)\Omega(x')}}\ \bohm_{ABCD}^{\text{\tiny{S}}}[x;x']\ H(v-v') \delta^{D-2}(\mathbf{x}-\mathbf{x}')\, .}
\end{equation}
By substituting the above {expression} in \eqref{Omega-inverse'}, we have
\begin{equation}\label{main-inverse-eq}
    \begin{split}
        2\, P^{AB}_{CD}\ \delta(v-v')=&\ \int \d v'' 
        \sqrt{\frac{\Omega(v)}{\Omega(v')}}\,  \partial_{v''}\left[ {\cal G} ^{ABEF}[v;v'']\,\bohm_{EFCD}^{\text{\tiny{S}}}[v'';v']\ H(v''-v') \right] \delta(v-v'') 
       \\
    =&\ \int \d v'' \bigg[2\,   {\cal G}_{{\text{\tiny{TF}}}}^{ABEF}(v)\,\bohm_{EFCD}^{\text{\tiny{S}}}[v;v]\, \delta(v-v') \,  \delta(v-v'') \\
    & \hspace{-3 cm} +\sqrt{\frac{\Omega(v)}{\Omega(v')}}\,  \Big(\partial_{v} {\cal G} ^{ABEF}[v'';v]\,\bohm_{EFCD}^{\text{\tiny{S}}}[v;v']\
    +{\cal G}_{{\text{\tiny{TF}}}}^{ABEF}(v)\,\partial_{v}\bohm_{EFCD}^{\text{\tiny{S}}}[v;v']\Big)\ H(v-v') \,  \delta(v-v'')\bigg] \\
        =& 2  \,  {\cal G}_{{\text{\tiny{TF}}}}^{ABEF}(v)\,\bohm_{EFCD}^{\text{\tiny{S}}}[v;v]
        \delta(v-v')  \\
        +&  \sqrt{\frac{\Omega(v)}{\Omega(v')}}\, \left[  {\cal G}_{{\text{\tiny{TF}}}}^{ABEF}(v)\,\partial_{v}\bohm_{EFCD}^{\text{\tiny{S}}}[v;v']
         + \bohm_{EFCD}^{\text{\tiny{S}}}[v;v'] \boldsymbol{{\cal A}}^{ABEF}(v)\right]
         H(v-v')\, ,
    \end{split}
\end{equation}
where
\begin{equation}\label{calA-exp}
    \begin{split}
     \boldsymbol{{\cal A}}^{ABCD}(v) &=  \int \d v''    \partial_{v}  {\cal G} ^{ABCD}[v'';v] \, \delta(v-v'') 
     \\
     &=\frac{1}{2}\,  \partial_v\mathcal{G} ^{ABCD}+{\frac{1}{D-2}\left({\cal N}^{CD}(v)\gamma^{AB}(v)-{\cal N}^{AB}(v)\gamma^{CD}(v)\right)}\, \\
     &=   -\frac12 \Bigl[\gamma^{AC}(v)\mathcal{N}^{BD}(v)+\gamma^{AD}(v)\mathcal{N}^{BC}(v)
         +\mathcal{N}^{AC}(v)\gamma^{BD}(v)+\mathcal{N}^{AD}(v)\gamma^{BC}(v)\\
         &\ -\frac{4}{D-2}\gamma^{AB}(v)\mathcal{N}^{CD}(v) \Bigr]\, .
    \end{split}
\end{equation}
Note that $ \gamma_{AB}(v) \boldsymbol{{\cal A}}^{ABCD}(v)=0,  \boldsymbol{{\cal A}}^{ABCD}(v)\gamma_{CD}(v)=-2{\cal N}^{AB}$.


Eq.\eqref{main-inverse-eq} yields the following equations:
\paragraph{Continuity condition at $\boldsymbol{v=v'}$.} From the coefficient of the delta function in \eqref{main-inverse-eq}  we {can} read the following algebraic equation:
\begin{equation}\label{continuity-con-bohm-s}
     {\cal G}^{ABEF}(v)\,\bohm_{EFCD}^{\text{\tiny{S}}}[v;v]=P^{AB}_{CD}(v)\, ,
\end{equation}
which yields
\begin{equation}\label{continuity-con-bohm-s'}
\bohm_{ABCD}^{\text{\tiny{S}}}[v;v]={\cal G}_{ABCD}(v)\, .
\end{equation}
\paragraph{Differential equation for $\boldsymbol{v\neq v'}$.} From the coefficient of the Heaviside function in \eqref{main-inverse-eq}  we {can} read the following first-order differential equation:
\begin{equation}
         {\cal G} ^{ABEF}(v)\,\partial_{v}\bohm_{EFCD}^{\text{\tiny{S}}}[v;v']
         + \boldsymbol{{\cal A}}^{ABEF}(v)\bohm_{EFCD}^{\text{\tiny{S}}}[v;v'] =0\, ,
\end{equation}
This equation may be written as
\begin{equation}
  \partial_{v}\left( {\cal G} ^{ABEF}(v)\,\bohm_{EFCD}^{\text{\tiny{S}}}[v;v']\right) + \left(\boldsymbol{{\cal A}}^{ABEF}(v)- \partial_{v}{\cal G} ^{ABEF}(v)    \right) \bohm_{EFCD}^{\text{\tiny{S}}}[v;v'] =0\, .
\end{equation}
Now by using the explicit expression of $\boldsymbol{{\cal A}}^{ABEF}(v)$ in equation \eqref{calA-exp}, we get
\begin{equation}
  \partial_{v}\left( {\cal G} ^{ABEF}(v)\,\bohm_{EFCD}^{\text{\tiny{S}}}[v;v']\right) - \left(\frac12 \partial_{v}{\cal G} ^{ABEF}(v) -\frac{1}{D-2}{\cal N}^{EF}(v)\gamma^{AB}(v)\right) \bohm_{EFCD}^{\text{\tiny{S}}}[v;v'] =0\, ,
\end{equation}
where we used  $\gamma^{EF}(v)\bohm_{EFCD}^{\text{\tiny{S}}}[v;v']=0$, leading to
\begin{equation}\label{X-AB-CD}
  \partial_{v} \boldsymbol{{\cal X}}^{AB}{}_{CD}[v;v'] - \boldsymbol{{\cal B}}^{AB}{}_{EF}(v){\cal X}^{EF}{}_{CD}[v;v']  =0\, ,
\end{equation}
with \footnote{Since $\gamma_{CD}(v){\cal X}^{Cd}{}_{EF}[v;v']=0$, we have dropped the terms proportional to $\gamma_{CD}$ in  $\boldsymbol{{\cal B}}^{AB}{}_{CD}(v)$.}
\begin{equation}\begin{split}
   \boldsymbol{{\cal X}}^{AB}{}_{CD}[v;v'] := {\cal G} ^{ABEF}(v)\,\bohm_{EFCD}^{\text{\tiny{S}}}[v;v']\, ,\qquad  
    \boldsymbol{{\cal B}}^{AB}{}_{CD}(v) =
    -2{\cal N}^{(A}_{(C}\delta^{B)}_{D)}\,.
\end{split}
\end{equation}
We note that $\boldsymbol{{\cal B}}^{AB}{}_{CD}(v)\gamma^{CD}(v)=\partial_v\gamma^{AB}, \gamma_{AB}(v)\boldsymbol{{\cal B}}^{AB}{}_{CD}(v)=-\partial_v\gamma_{CD}$. 

{We solve this equation with the following continuity condition \eqref{continuity-con-bohm-s}, namely
\begin{equation}\label{continuity-cond-cal X}
    \mathcal{X}^{AB}{}_{CD}[v;v]=P^{AB}_{CD}[v]\, .
\end{equation}}
To solve \eqref{X-AB-CD} we introduce the evolution matrix for $v\geq v'$
\begin{equation}\label{U-evolution}
   \boxed{ \boldsymbol{{\cal U}}^{AB}{}_{CD}[v;v'] = \mathbf{V} \exp{\left[\int^{v}_{v'} \d{}\tilde{v}\ \boldsymbol{{\cal B}}(\tilde{v})\right]}^{AB}_{ \ CD}\, , \qquad \boldsymbol{{\cal U}}^{AB}{}_{CD}[v;v]=\delta^{(A}_{(C}\delta^{B)}_{D)}\, \ }
\end{equation}
where $\mathbf{V}$ denotes $v$-ordering and recalling the Dyson series, we have 
\begin{equation}
    \partial_v \boldsymbol{{\cal U}}^{AB}{}_{CD}[v;v']= \boldsymbol{{\cal B}}^{AB}{}_{EF}(v) \boldsymbol{{\cal U}}^{EF}{}_{CD}[v;v'] \qquad v\geq v'\, .
\end{equation}
So, imposing \eqref{continuity-cond-cal X} to fix the integration constant, we learn
\begin{equation}
   {\cal X}^{AB}{}_{CD}[v;v']= \boldsymbol{{\cal U}}^{AB}{}_{EF}[v;v'] P^{EF}_{CD}(v') \qquad v\geq v'\, .
\end{equation}
Noting that ${\cal G} ^{ABEF}(v) {\cal G}_{EFCD}(v)=P_{AB}^{CD}(v)$ and that $P_{AB}^{EF}(v)\bohm_{EFCD}^{\text{\tiny{S}}}[v;v']=\bohm_{ABCD}^{\text{\tiny{S}}}[v;v']$, we have
\begin{equation}
    \bohm_{ABCD}^{\text{\tiny{S}}}[v;v']= {\cal G}_{ABEF}(v){\cal X}^{EF}_{CD}[v;v'] =
    {\cal G}_{ABEF}(v)\boldsymbol{{\cal U}}^{EF}{}_{GH}[v;v'] P^{GH}_{CD}(v')\, , \qquad v\geq v'\, .
\end{equation}
The solution for $v'\geq v$ is then fixed through the symmetry requirement, $\bohm_{ABCD}^{\text{\tiny{S}}}[v;v']=\bohm_{CDAB}^{\text{\tiny{S}}}[v';v]$. Explicitly, 
\begin{equation}\label{bohm-vv'-final}
   \boxed{ \bohm_{ABCD}^{\text{\tiny{S}}}[v;v']=\left\{\begin{array}{cc}
    {\cal G}_{ABEF}(v)\ \boldsymbol{{\cal U}}^{EF}{}_{GH}[v;v'] P^{GH}_{CD}(v')\, , \qquad  & \qquad v\geq v' \\ \, \\ 
    {\cal G}_{CDEF}(v')\ \boldsymbol{{\cal U}}^{EF}{}_{GH}[v';v] P^{GH}_{AB}(v)\, , \qquad  & \qquad v'\geq v\end{array}\right.}
\end{equation}
\paragraph{Discussion.} Given the above result, which {may} look complicated, some clarifying comments are in order:
\begin{enumerate}
    \item One may explicitly verify that
    \begin{equation}
       \hspace{-0.5 cm} \gamma^{AB}(v)\bohm_{ABCD}^{\text{\tiny{S}}}[v;v']=0\, , \qquad \bohm_{ABCD}^{\text{\tiny{S}}}[v;v']\gamma_{CD}(v')=0\, , \qquad
\bohm_{ABCD}^{\text{\tiny{S}}}[v;v']=\bohm_{CDAB}^{\text{\tiny{S}}}[v';v] \, ,
    \end{equation}
and also
\begin{equation}
    P^{AB}_{EF}(v)\bohm_{ABCD}^{\text{\tiny{S}}}[v;v']=\bohm_{EFCD}^{\text{\tiny{S}}}[v;v'] \, , \qquad P^{CD}_{EF}(v')\bohm_{ABCD}^{\text{\tiny{S}}}[v;v']=\bohm_{ABEF}^{\text{\tiny{S}}}[v;v'] \, .
\end{equation}
\item As \eqref{main-inverse-eq} shows $\bohm_{ABCD}^{\text{\tiny{S}}}[v;v']$ is  a smooth function in $v,v'$ and in particular $\bohm_{ABCD}^{\text{\tiny{S}}}[v;v]={\cal G}_{ABCD}(v)$.
\item Recalling that $2{\cal N}_{AB}(v)=\partial_v\gamma_{AB}$, $\boldsymbol{{\cal U}}^{AB}{}_{CD}[v;v']$  evolves a given configuration of gravitational waves, parametrized in $\gamma_{AB}$, from an arbitrary boundary $v'$ to $v$ (for $v\geq v'$).
\item The propagator $\boldsymbol{{\cal U}}^{AB}{}_{CD}[v;v']$ only changes the $v$ dependence of the gravitational waves and {does} not {affect}  $\mathbf{x}$. This is expected as two points $(v,\mathbf{x})$ and $(v',\mathbf{x}')$ on ${\cal N}$ are  causally connected only when $\mathbf{x}=\mathbf{x}'$.
\item {The computation of bracket can be depicted as shown in Fig.\ref{Fig:vv'-evolution}. }  To find the bracket between the two generic points $v,v'$ we need to start {by}  {evolving} the one at $v'<v$ to the one at $v$ and then compare the two. The bracket is  obtained {by} noting that 
$\{\gamma_{AB}(v), \gamma_{CD}(v)\}=0$ and {using} the definition of projector $P^{AB}_{CD}(v)$.
\item One also {can} verify that 
\begin{equation}\label{gamma-PB-appen-UD}
    \left\{ \gamma_{AB}(x) ,  \gamma^{CD}(x')\right\}= \frac{16\pi G}{\sqrt{\Omega(x)\Omega(x')}}\ \bohm_{ABEF}^{\text{\tiny{S}}}[x;x']\ {\cal G}^{EFCD}(x')\ H(v-v') \delta^{D-2}(\mathbf{x}-\mathbf{x}')\, .
\end{equation}

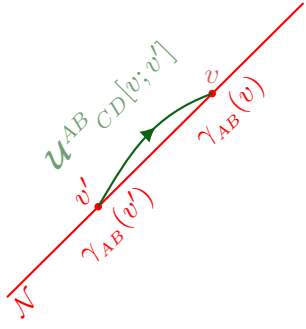
\begin{figure}[t]
\def \L {3.0}
    \centering
\begin{tikzpicture}
  \draw[thick,red] (-0.2*\L,-0.2*\L) coordinate (b) -- (1.1*\L,1.1*\L) coordinate (t);
  \draw[red] (-0.05*\L,-0.15*\L) node[left, rotate=45] (scrip) {\small{${\cal N}$}};
  \node (v1) at (0.2*\L,0.2*\L) {};
  \fill[red] (v1) circle (0.05) node[above, rotate=45] (scrip) {$v^\prime$} node[below, rotate=45] (scrip) {$\gamma_{_{AB}}(v^\prime)$};
  \node (v2) at (0.7*\L,0.7*\L) {};
  \fill[red] (v2) circle (0.05) node [above,darkred!60] {$v$} node[below, rotate=45] (scrip) {$\gamma_{_{AB}}(v)$};
  \draw[green, mid arrow={>}, thick] ($(v1)+(0*\L,0*0.05*\L)$) 
        to[out=60,in=200, looseness=1] ($(v2)+(-0*0.05*\L,0.0*\L)$) ;
  \node (v3) at (0.45*\L,0.45*\L) {};
  \fill (v3)+(-0.2*\L,0.2*\L)  node [green!60,rotate=45] {$\boldsymbol{{\cal U}}^{AB}{}_{CD}[v;v']$};
\end{tikzpicture}
\caption{Depiction of two points $v'<v$ located at the same position transverse $\mathbf{x}=\mathbf{x}'$. A configuration of gravitational waves is evolved from $v'$ to $v$ using the evolution matrix $\boldsymbol{{\cal U}}^{AB}{}_{CD}[v;v']$. This evolution matrix do not change $\mathbf{x}$.
}\label{Fig:vv'-evolution}
\end{figure}

\item Recalling that on-shell ${\cal N}_{AB}=\frac12\partial_v\gamma_{AB}$ and using \eqref{gamma-PB-appen}, one can compute $\{{\cal N}_{AB}(v), \gamma_{CD}(v')\}$ on-shell {as follows}:
\begin{equation}
\begin{split}
    \{{\cal N}_{AB}(x), \gamma_{CD}(x')\}&=\Bigg[\frac{16\pi G}{\Omega(x)} {\cal G}_{ABCD}(x) \delta(v-v')  +{4\pi G}{\sqrt{\frac{\Omega(x)}{\Omega(x')}}}\bigg(\partial_v(\frac{1}{\Omega}{\cal G}_{ABEF}(x))\\ & -\frac{2}{(D-2)\Omega}\gamma_{AB}(x){\cal N}_{EF}(x)\bigg) \boldsymbol{{\cal U}}^{EF}{}_{KL}[v;v'] P^{KL}_{CD}(v') H(v-v') \Bigg]  \delta^{D-2}(\mathbf{x}-\mathbf{x}')\, .
\end{split}
\end{equation}
The  $\delta(v-v')$ term in the above is what we had in the off-shell expression \eqref{PB-NN''}. The $H(v-v')$ terms, however, appear on-shell, as the evolution operator of the system and the WdW metric are both $\gamma_{AB}$ dependent.
\item \textbf{Jacobi Identity.} We obtained the brackets through inverting the symplectic 2-form \eqref{Bulk-I-1} in {the} $\delta\Omega=0$ sector, which is clearly closed, {as indicated by}  the first line {in} \eqref{Bulk-I-1}. {The} closedness of $\bO^{\text{\tiny{I}}}_{\text{\tiny Bulk}}$ implies {the} Jacobi identity for the bracket \eqref{gamma-PB-appen} obtained from inverting the symplectic form. Here we demonstrate this well-known statement for our specific case. Let us start from \eqref{Omega-inverse'}, {and by} taking derivatives w.r.t to {the} ``trace-free'' $\hat\gamma_{KL}(y)$, i.e.  $\frac{\delta\gamma_{EF}}{\delta\hat\gamma_{KL}(y)}=P_{EF}^{KL}$, we have:
\begin{equation}\label{Jacobi-1}
     \int^{''}_{\cal N} \ \frac{\delta\bO^{ABEF}[x;x'']}{\delta \hat\gamma_{KL}(y)}\bohm_{EFCD}[x'';x']+\bO^{ABEF}[x;x'']\frac{\delta\bohm_{EFCD}[x'';x']}{\delta \hat\gamma_{KL}(y)}=0
\end{equation}
We may then use closedness relation,
\begin{equation}
    \frac{\delta\bO^{ABEF}[x;x'']}{\delta \hat\gamma_{KL}(y)}+\frac{\delta\bO^{KLAB}[y;x]}{\delta \hat\gamma_{EF}(x'')}+\frac{\delta\bO^{EFKL}[x'';y]}{\delta \hat\gamma_{AB}(x)}=0 \, , 
\end{equation}
\eqref{Omega-inverse'} and {through} straightforward algebraic manipulations using symmetry properties of $\bohm_{ABCD}[x;x']$  we arrive at: 
\begin{equation}
     \int^{y}_{\cal N} \ \bohm_{KLEF}[y;x'']\ \frac{\delta\bohm_{ABCD}[x;x']}{\delta \hat\gamma_{KL}(y)}+ \bohm_{KLCD}[y;x']\ \frac{\delta\bohm_{EFAB}[x'';x]}{\delta \hat\gamma_{KL}(y)}+\bohm_{KLAB}[y;x]\ \frac{\delta\bohm_{CDEF}[x';x'']}{\delta \hat\gamma_{KL}(y)}=0
\end{equation}
which is immediately resulting in 
\begin{equation}
     \begin{split}
         &\left\{\left\{ \gamma_{AB}(x_{_1}) ,  \gamma_{CD}(x_{_2})\right\}, \gamma_{EF}(x_{_3})\right\}+ \left\{\left\{ \gamma_{EF}(x_{_3}) ,  \gamma_{AB}(x_{_1})\right\}, \gamma_{CD}(x_{_2})\right\} \\
         &+ \left\{\left\{ \gamma_{CD}(x_{_2}) ,  \gamma_{EF}(x_{_3})\right\}, \gamma_{AB}(x_{_1})\right\}=0
     \end{split}
\end{equation}
\end{enumerate}
\section{ On-shell symplectic form, \texorpdfstring{$D=4$}{} case}\label{appen-4d}
Here, we work out the physically interesting case of $D=4$. The on-shell solution space is governed by four boundary modes $\Omega(v_b,\mathbf{x}), {\cal P}(v_b, \mathbf{x})$ and ${\cal S}_A(v_b, \mathbf{x})$ $(A=1,2)$ and three bulk modes $\Omega(v,\mathbf{x}), \gamma_{AB}(v,\mathbf{x})$.  In this case, $\gamma_{AB}$ has only two independent d.o.f which may be parametrized as a $2\times 2$ matrix
\begin{equation}\label{gamma-01}
\gamma_{AB}=\begin{pmatrix} e^{\gamma}\cosh\phi & \sinh\phi 
\\ \sinh\phi & e^{-\gamma}\cosh\phi\end{pmatrix}\, , \qquad \gamma^{AB}=\begin{pmatrix} e^{-\gamma} \cosh\phi & -\sinh\phi \\ -\sinh\phi & e^{\gamma}\cosh\phi\end{pmatrix}\, .
\end{equation}
The news tensor readily can be read as follows
\begin{equation}
\mathcal{N}_{AB}=\frac{1}{2}\partial_{v}{\gamma}
\begin{pmatrix} 
e^{\gamma} \cosh\phi & 0 \\
0 & -e^{-\gamma} \cosh\phi
\end{pmatrix}
+\frac{1}{2}\partial_{v}{\phi}\begin{pmatrix} e^{\gamma}\sinh\phi & \cosh\phi \\ \cosh\phi & e^{-\gamma}\sinh\phi\end{pmatrix}\, ,
\end{equation}
and also with upper indices as 
\begin{equation}
\mathcal{N}^{AB}=\frac{1}{2}\,  \partial_{v}\gamma
\begin{pmatrix}
e^{-\gamma} \cosh\phi & 0
\\ 
0 & -e^{\gamma} \cosh\phi
\end{pmatrix}
+\frac{1}{2}\partial_{v}\phi\begin{pmatrix} -e^{-\gamma}\sinh\phi & \cosh\phi \\ \cosh\phi & -e^{\gamma}\sinh\phi\end{pmatrix}\, .
\end{equation}
resulting in
\begin{equation}\label{N2-4d}
 {\cal N}^2=\frac{1}{2}(\partial_{v}\gamma)^2\cosh^2{\phi}+\frac{1}{2}(\partial_{v}\phi)^2\, .
\end{equation}
\paragraph{WdW metric.} The WdW metric, 
\begin{equation}
    \begin{split}
        {\cal G}^{ABCD}&=\frac12(\gamma^{AC}\gamma^{BD}+\gamma^{AD}\gamma^{BC}-\gamma^{AB}{\gamma^{CD}})\, , \\
     {\cal G}_{ABCD}&=\frac12(\gamma_{AC}\gamma_{BD}+\gamma_{AD}\gamma_{BC}-\gamma_{AB}{\gamma_{CD}})\, , 
    \end{split}
\end{equation}
has the following components
\begin{equation}\label{4d-WdW-metric}
\begin{split}
    {\cal G}^{1111}= \frac12 e^{-2\gamma}\cosh^2\phi\, , &\qquad {\cal G}^{2222}= \frac12 e^{2\gamma}\cosh^2\phi\, ,\\ 
    {\cal G}^{1122}={\cal G}^{2211}=-\frac12(1-\sinh^2\phi)\, ,&\qquad {\cal G}^{1212}={\cal G}^{2121}=\frac12(1+\sinh^2\phi)\, ,\\
    {\cal G}^{1112}={\cal G}^{1211}=-\frac14e^{-\gamma}\sinh2\phi\, ,&\qquad {\cal G}^{1222}={\cal G}^{2212}=-\frac14e^{\gamma}\sinh2\phi\, .
\end{split}
\end{equation}
With the above, $\slashed{\delta}\hat{\gamma}_{AB}$ \eqref{delta-slashes} is obtained as
\begin{equation}
\slashed{\delta}\hat{\gamma}_{AB}=\begin{pmatrix} e^{\gamma}(\cosh\phi\, \slashed\delta \hat{\gamma}+ \sinh \phi\, \slashed\delta \hat{\phi}) & \cosh\phi\, \slashed\delta \hat{\phi}
\\ \cosh\phi\, \slashed\delta \hat{\phi} & e^{-\gamma}(-\cosh\phi\, \slashed\delta \hat{\gamma}+\sinh \phi\, \slashed\delta \hat{\phi})\end{pmatrix}\, ,
\end{equation}
where we introduced non-exact forms on solution space as follows
\begin{equation}
    \begin{split}
         \slashed\delta \hat{\mathcal{P}}=\delta {\mathcal{P}}-\Omega \frac{\partial_{v}{\phi}}{\partial_{v}{\Omega}}&\delta\phi-\Omega\cosh^{2} \phi \frac{\partial_{v}{\gamma}}{\partial_{v}{\Omega}}\delta \gamma\, ,\\
         \slashed\delta \hat{\gamma}= \delta \gamma - \frac{\partial_{v}{\gamma}}{\partial_{v}{\Omega}} \delta \Omega\,, &\qquad 
        \slashed\delta \hat{\phi}= \delta \phi -\frac{\partial_{v}{\phi}}{\partial_{v}{\Omega}}\delta\Omega\, .
    \end{split}
\end{equation}
As a next step, we compute the symplectic potential \eqref{Theta-on-shell} and symplectic form \eqref{Omega-on-shell}
\begin{subequations}
    \begin{align}
        & 16\pi G \ \bTh_{\text{\tiny on-shell}} =  \oint_{\mathcal{N}_b}\, \Omega \, \delta \mathcal{P} - \int_{\mathcal{N}}  \Omega\left[\partial_{v}{\phi}\, \slashed\delta \hat{\phi}+\cosh^{2}\phi\, \partial_v{\gamma} \,\slashed\delta \hat{\gamma}\right]\, , \\
        &16\pi G \ \bO_{\text{\tiny on-shell}}= \oint_{\mathcal{N}_b} \delta \Omega \wedge \delta \mathcal{P}   +\int_{\mathcal{N}} \Omega\left[ \sinh 2\phi \, \partial_{v}\gamma\, \slashed{\delta} \hat{\gamma}\wedge \slashed{\delta} \hat{\phi}+\cosh^{2}\phi\, \slashed{\delta} \hat{\gamma}\wedge \partial_{v}\slashed{\delta} \hat{\gamma}+\slashed{\delta} \hat{\phi} \wedge \partial_{v} \slashed{\delta} \hat{\phi} \right]\, .
    \end{align}
\end{subequations}
Neither {the} bulk nor {the} boundary parts are closed or invertible. {The} non-invertibility of the boundary part is due to its non-closedness, which is a consequence of having the boundary symplectic flux \eqref{F-bdry}
\begin{equation}
    \mathbf{F}_{\text{\tiny bdy}}=\frac{1}{16\pi G}\oint_{{\cal N}_b} \frac{\Omega}{\partial_{v}{\Omega}}\left(\partial_{v}{\phi}\delta\phi-\cosh^2\phi \partial_{v}{\gamma}\delta \gamma\right)\wedge \delta \Omega\, .
\end{equation}
Non-invertibility of the bulk piece is, however, due to both the presence of the flux (manifested in non-closed 1-forms $\slashed{\delta}X$) and the existence of a  kernel vector in solution space (the Carrollian nature of the bulk solution space geometry). 

\paragraph{Carrollian bulk solution space.} Let us rewrite the bulk term in the symplectic form as
\begin{equation}
    \ \bO_{\text{\tiny Bulk}}= \frac{1}{32\pi G } \, \int_{\mathcal{N}}\, \int^{\prime}_{\mathcal{N}} \, \delta \varphi^{\mathbb{I}}(v,\mathbf{x}) \, \bO_{_{\mathbb{IJ}}}[v,\mathbf{x};v',\mathbf{x}']\wedge \delta \varphi^{\mathbb{J}}(v',\mathbf{x}')\, ,
\end{equation}
where $\varphi^{\mathbb{I}}=\{\Omega, {\gamma}, \phi\}$ and as in section \ref{sec:soln-space-carrollian} we define the bulk solution space metric as in \eqref{bulk-soln-metric}
\begin{equation}\label{4d-bulk-Omega-1}
    \bO_{_{\mathbb{IJ}}}[x;x']= \bG_{_{\mathbb{IJ}}}[x;x'] \, \partial_{v}\delta(v-v') \delta^{D-2}(\mathbf{x}-\mathbf{x}')\, .
\end{equation}
The $3\times3$ function-valued matrix $\bG_{_{\mathbb{IJ}}}[x;x']$ leading to on-shell bulk solution space metric 
\begin{equation}\label{4d-METRIC}
\bG_{_{\mathbb{IJ}}}=2\Omega \ 
\begin{pmatrix}
   2\frac{{\cal N}^2}{(\partial_v\Omega)^2} & -\frac{\partial_v\gamma}{\partial_v\Omega}\cosh^2\phi& -\frac{\partial_v\phi}{\partial_v\Omega}
\\ 
 -\frac{\partial_v\gamma}{\partial_v\Omega}\cosh^2\phi & \cosh^2\phi & 0\\ 
-\frac{\partial_v\phi}{\partial_v\Omega} & 0 & 1
\end{pmatrix}\, ,  
\end{equation}
where ${\cal N}^2$ is given in \eqref{N2-4d}. More explicitly (cf. \eqref{Bulk-METRIC}),
\begin{equation}\label{Bulk-METRIC-4d}
\begin{split}
    \delta{\mathbf{\mathbb{S}}}^2 &=2\int_{\cal N} \int^{'}_{\cal N} \sqrt{\Omega(x)\Omega(x')}\ \biggl\{\cosh({\gamma(x)-\gamma(x')})\ \slashed{\delta}\hat{\phi}(x)\slashed{\delta}\hat{\phi}(x') \\ &  +
    \left(\cosh \phi(x)\, \cosh \phi(x')-\cosh(\gamma(x)-\gamma(x'))\sinh \phi(x)\, \sinh \phi(x')\right)\ \slashed{\delta}\hat{\gamma}(x)\slashed{\delta}\hat{\gamma}(x')\\
    &+\frac{1}{2}\sinh({\gamma(x)-\gamma(x')})\left(\sinh 2 \phi(x')\ \slashed{\delta}\hat{\phi}(x)\slashed{\delta}\hat{\gamma}(x'){-}\sinh 2 \phi(x)\ \slashed{\delta}\hat{\phi}(x')\slashed{\delta}\hat{\gamma}(x)\right) 
    \biggr\}\, .
    \end{split}
\end{equation}
The above metric is clearly defining a Carrollian geometry, with a $2\times 2$ ``Carrollian metric'' 
\begin{equation}\label{4d-2x2}
\bG_{ij}=2\Omega \ 
\begin{pmatrix}
 \cosh^2\phi & 0\\ 
 0 & 1
\end{pmatrix}\, , 
\end{equation}
and the kernel vector $\mathbf{K}$ and the Ehresmann connection $\mathbf{E}$,
\begin{equation}
     \mathbf{K}=\mathbf{K}^{^\mathbb{I}}\frac{\delta}{\delta\phi^{\mathbb{I}}}= \partial_{v}\Omega \frac{\delta}{\delta\Omega}+ \partial_{v}\phi \frac{\delta}{\delta\phi}+\partial_{v}\gamma\frac{\delta}{\delta\gamma}\, ,\qquad \mathbf{E}=\mathbf{E}_{_{\mathbb{I}}}\delta\phi^{\mathbb{I}}= \frac{1}{ \partial_v \Omega}\, \delta \Omega+\frac{1}{ \partial_v \gamma}\, \delta \gamma+\frac{1}{ \partial_v \phi}\, \delta \phi\, .
\end{equation}
As we see $\bG_{_{\mathbb{IJ}}}\mathbf{K}^{^\mathbb{J}}=0$ and $\mathbf{K}^{^\mathbb{I}}\mathbf{E}_{_{\mathbb{I}}}=1$. 
\paragraph{4d bulk on-shell Poisson brackets.} 
In our specific parametrization for $\gamma_{AB}$ in terms of $\gamma$ and $\phi$, one can make the analysis of appendix \ref{appen-inverting} more explicit and find on-shell Poisson brackets of these variables. To this end, and a check of our previous analysis, we repeat them in this parametrization. 
One can simply read the integrable part of the symplectic form \eqref{Integ-Omega} (i.e. in the $\delta\Omega=0$ sector) 
\begin{equation}\label{Bulk-I-2-4D}
    \begin{split}
         \bO^{\text{\tiny{I}}}_{\text{\tiny Bulk}} &  = \frac{1}{16 \pi G}\int_{\mathcal{N}} \Omega\,   \delta\mathcal{N}_{AB} \wedge {\delta}{\gamma}^{AB} \\
         & =\frac{1}{16 \pi G}\int_{\mathcal{N}} \Omega\left[ \sinh 2\phi \, \partial_{v}\gamma\, \delta {\gamma}\wedge \delta {\phi}+\cosh^{2}\phi\, \delta {\gamma}\wedge \partial_{v}\delta {\gamma}+\delta {\phi} \wedge \partial_{v}\delta {\phi} \right]\, .
        \end{split}
\end{equation}
As a check, one may immediately observe that $\bO^{\text{\tiny{I}}}_{\text{\tiny Bulk}}$ is closed in $\delta\Omega=0$ sector, i.e. $\delta \bO^{\text{\tiny{I}}}_{\text{\tiny Bulk}}=0$. This symplectic form  may be written as
\begin{equation}
    16\pi G \ \bO^{\text{\tiny{I}}}_{\text{\tiny Bulk}}= \frac{1}{2} \, \int_{\mathcal{N}}\, \int^{\prime}_{\mathcal{N}} \, \delta \varphi^{i}(v,\mathbf{x}) \, \bO_{ij}[v,\mathbf{x};v',\mathbf{x}']\wedge \delta \varphi^{j}(v',\mathbf{x}')\, ,
\end{equation}
where $\varphi^{i}=\{{\gamma}, \phi\}$. By comparing the last two equations, one can simply find 
\begin{equation}\label{D01}
    \bO_{ij}[x;x']= -\mathbf{A}_{ij}[x;x'] \, \partial_{v'}\delta(v-v')\delta^{2}(\mathbf{x}-\mathbf{x}') +\mathbf{B}_{ij}[x;x'] \, \delta(v-v')\delta^{2}(\mathbf{x}-\mathbf{x}')\, ,
\end{equation}
with
\begin{equation}
    \mathbf{A}_{ij}[x;x']= \hat{\mathbf{A}}_{ij}[x]+\hat{\mathbf{A}}_{ji}[x'] \, , \qquad \mathbf{B}_{ij}[x;x']= \hat{\mathbf{B}}_{ij}[x]-\hat{\mathbf{B}}_{ji}[x'] \, ,
\end{equation}
where the explicit form of these matrices is given by
\begin{equation}
  \hat{\mathbf{A}}_{ij}[x]= \Omega
\begin{pmatrix}
 \cosh^2\phi & 0\\ 
 0 & 1
\end{pmatrix}\, , \qquad 
\hat{\mathbf{B}}_{ij}[x]=\Omega\, \sinh2\phi \, \partial_v\gamma \, \begin{pmatrix}
 0 & 1  \\
 0 & 0
\end{pmatrix} \, .
\end{equation}
The next step is to compute the inverse of the on-shell symplectic 2-form
\begin{equation}\label{Def-Inverse}
     \int_{\mathcal{N}}^{''}\, \bO_{ik}[v,\mathbf{x};v'',\mathbf{x}'']\, \bohm^{kj}[v'',\mathbf{x}'';v',\mathbf{x}']=\delta_{i}^{j}\, \delta(v-v')\delta^{2}(\mathbf{x}-\mathbf{x}')\, .
\end{equation}
This equation yields
\begin{equation}
    - \int_{\mathcal{N}}^{''} \mathbf{A}_{ik}[x;x'']  \, \bohm^{kj}[x'';x']\, \partial_{v''}\delta(v-v'')+\mathbf{B}_{ik}[x;x] \, \bohm^{kj}[x;x']=\delta_{i}^{j}\, \delta(v-v')\delta^{2}(\mathbf{x}-\mathbf{x}')\, .
\end{equation}
One can simplify this equation further
\begin{equation}
    \begin{split}
    &\hat{\mathbf{A}}_{ik}[x]\partial_{v}\bohm^{kj}[x;x']+\partial_{v}\left(\hat{\mathbf{A}}_{ki}[x]  \, \bohm^{kj}[x;x']\, \right)+\mathbf{B}_{ik}[x] \, \bohm^{kj}[x;x']=\delta_{i}^{j}\, \delta(v-v')\delta^{2}(\mathbf{x}-\mathbf{x}')\, .
    \end{split}
\end{equation}
One can rewrite this equation as follows
\begin{equation}\label{inverse-eq-4}
   {\mathbf{A}}_{ik}[x]\partial_{v}\bohm^{kj}[x;x']+\mathbf{C}_{ik}[x] \, \bohm^{kj}[x;x']=\delta_{i}^{j}\, \delta(v-v')\delta^{2}(\mathbf{x}-\mathbf{x}')\, ,
\end{equation}
where
\begin{equation}
    {\mathbf{A}}_{ik}[x]:={\mathbf{A}}_{ik}[x;x]
    \, , \quad   {\mathbf{B}}_{ik}[x]:={\mathbf{B}}_{ik}[x;x]\, , \quad \mathbf{C}_{ik}[x]:=\mathbf{B}_{ik}[x]+\partial_{v}\hat{\mathbf{A}}_{ki}[x]\, .
\end{equation}
Their explicit forms are given by
\begin{equation}
  {\mathbf{A}}_{ij}[x]= 2\, \Omega
\begin{pmatrix}
 \cosh^2\phi & 0\\ 
 0 & 1
\end{pmatrix} \, , \qquad 
{{\mathbf{C}}_{ij}[x]=\Omega\, \sinh{2\phi}\, \begin{pmatrix}
\partial_v\phi   &   \partial_v\gamma  \\
 -\partial_v\gamma & 0
\end{pmatrix}  
+ \frac{1}{2}\, \frac{\partial_v \Omega}{\Omega}\,  {\mathbf{A}}_{ij}[x]\, , }
\end{equation}
To solve \eqref{inverse-eq-4} we take the following ansatz
\begin{equation}
    \bohm^{ij}[x;x']={\frac{1}{\sqrt{\Omega(x)\Omega(x')}}}  \, \bohm_{\text{\tiny S}}^{ij}[v,v';\mathbf{x}] H(v-v')\delta^{2}(\mathbf{x}-\mathbf{x}')\, .
\end{equation}
To avoid cluttering and ease of notation, hereafter we will suppress the $\mathbf{x}$ dependence of our quantities e.g. $\bohm_{\text{\tiny S}}^{ij}[v,v';\mathbf{x}]=\bohm_{\text{\tiny S}}^{ij}[v,v']$.  By using this ansatz \eqref{inverse-eq-4} yields
\begin{equation}\label{discontinuity-con-1}
     \bohm_{\text{\tiny S}}^{ij}[v;v]=\frac{1}{2}\, \Omega\, {\mathbf{A}}^{ij}[v]=\frac{1}{4\, \cosh^2{\phi}}\,  
     \begin{pmatrix}
         1 & 0 \\
         0 & \cosh^2{\phi}
     \end{pmatrix}\, ,
\end{equation}
and
\begin{equation}
    \partial_v\bohm^{ij}_{\text{\tiny S}}[v;v']+ \mathbf{D}^{i}{}_{k}\, \bohm_{\text{\tiny S}}^{kj}[v;v'] =0\, ,
\end{equation}
where
\begin{equation}
    \mathbf{D}^{i}{}_{j}=\mathbf{A}^{ik} \mathbf{C}_{kj} -\frac{1}{2}\, \frac{\partial_v \Omega}{\Omega} \delta^{i}_{j}= \tanh{\phi} 
    \begin{pmatrix}
        \partial_v \phi & \partial_v \gamma \\
        -\partial_v \gamma\, \cosh^2{\phi} & 0
    \end{pmatrix}\, .
\end{equation}
Now let us consider
\begin{equation}
    \bohm^{ij}_{\text{\tiny S}}[v;v']= 
    \begin{pmatrix}
        X(v,v') & Y(v,v') \\
        Y(v',v) & Z(v,v')
    \end{pmatrix}\, ,
\end{equation}
where $X(v,v')=X(v',v)$ and $Z(v,v')=Z(v',v)$, then we need to solve four equations
\begin{subequations}
    \begin{align}
    & \partial_v \left[\cosh{\phi}(v) \, X(v,v')\right] +Y(v',v) \,\sinh{\phi}(v) \,  \partial_v \gamma(v) =0\, ,\label{G-a}\\
    & \partial_v Z(v,v') -\frac{1}{2 } Y(v,v') \, \sinh{2\phi(v)}\, \partial_v \gamma(v) =0\, ,\label{G-b}\\
    & \partial_v [\cosh{\phi}(x)\, Y(v,v')]  +Z(v,v') \, \sinh{\phi(v)}\,  \partial_v \gamma(v)=0\, , \label{G-c}\\
    & \partial_v Y(v',v)-\frac{1}{2 } X(v,v') \, \sinh{2\phi(v)}\, \partial_v \gamma(v)=0\, .\label{G-d}
    \end{align}
\end{subequations}
To solve these equations, let us do a change of variables as
\begin{equation}
    Z(v,v')=\frac{1}{4}\,  \cos{\chi(v,v')}\, , \qquad Y(v,v')= \frac{1}{4}\, \frac{\sin{\chi(v,v')}}{\cosh \phi(v)}\, .
\end{equation}
Using the above, equations \eqref{G-b} and \eqref{G-c} become a single equation
\begin{equation}\label{chi-eq}
    \partial_{v}\chi(v,v') + \sinh \phi(v) \partial_{v}\gamma(x) =0\, ,
\end{equation}
which can be solved as follows
\begin{equation}
     \chi(v,v')=-\chi(v',v)=-\int^{v}_{v'} \d{} \tilde{v} \, \sinh \phi(\tilde{v}) \partial_{\tilde{v}}\gamma(\tilde{v}) \, .
\end{equation}
Next, we solve \eqref{G-d} and find that 
\begin{equation}
    X(v,v')=\frac{\cos \chi(v,v')}{4 \cosh \phi (v)\, \cosh \phi(v')}\, .
\end{equation}
One can readily check that the initial condition \eqref{discontinuity-con-1} is satisfied.
Now Dirac brackets read as
\begin{subequations}\label{001}
    \begin{align}
        &\{\gamma(v), \gamma(v')\}={\frac{{8\pi G}}{\sqrt{\Omega(x)\Omega(x')}}} \frac{\cos\chi(v,v')}{\cosh \phi (x)\, \cosh \phi(x')} \,  H(v-v')\delta^{2}(\mathbf{x}-\mathbf{x}') \, , \\
        &\{\gamma(x), \phi(x')\}=\frac{{8\pi G}}{\sqrt{\Omega(x)\Omega(x')}}\, \frac{\sin\chi(v,v')}{\cosh \phi (x)} \,  H(v-v')\delta^{2}(\mathbf{x}-\mathbf{x}') \, ,\\
        &\{\phi(x), \gamma(x')\}=-\frac{{8\pi G}}{\sqrt{\Omega(x)\Omega(x')}}\, \frac{\sin\chi(v,v')}{\cosh \phi (x')} \,  H(v-v')\delta^{2}(\mathbf{x}-\mathbf{x}') \, ,\\
        &\{\phi(x), \phi(x')\}= \frac{{8\pi G}}{\sqrt{\Omega(x)\Omega(x')}}\, \cos \chi(v,v') \,  H(v-v')\delta^{2}(\mathbf{x}-\mathbf{x}') \, ,
        \end{align}
\end{subequations}
where $\chi(v,v')$ is given by \eqref{chi-eq}.
Therefore
\begin{subequations}
    \begin{align}
        & \hspace{-0.9 cm}\bohm_{1111}^{\text{\tiny{S}}}[x;x']=\frac{1}{2}\, \sqrt{\frac{\gamma_{11}(x)}{\gamma_{22}(x)}}\sqrt{\frac{\gamma_{11}(x')}{\gamma_{22}(x')}}\left[(\gamma_{12}(x')-\gamma_{12}(x))\sin\chi(v,v')+(\gamma_{12}(x)\gamma_{12}(x')+1)\cos\chi(v,v')\right]\, ,\\
        & \hspace{-0.9 cm}\bohm_{1112}^{\text{\tiny{S}}}[x;x']=\frac{1}{2}\,\sqrt{\frac{\gamma_{11}(x)}{\gamma_{22}(x)}}\sqrt{{\gamma_{11}(x')}{\gamma_{22}(x')}}\ \left[\sin\chi(v,v')+\gamma_{12}(x)\cos\chi(v,v')\right]\, ,\\
        &\hspace{-0.9 cm} \bohm_{1122}^{\text{\tiny{S}}}[x;x']=\frac{1}{2}\,\sqrt{\frac{\gamma_{11}(x)}{\gamma_{22}(x)}}\sqrt{\frac{\gamma_{22}(x')}{\gamma_{11}(x')}} \left[(\gamma_{12}(x')+\gamma_{12}(x))\sin\chi(v,v')+(\gamma_{12}(x)\gamma_{12}(x')-1)\cos\chi(v,v')\right]\, ,\\
        &\hspace{-0.9 cm} \bohm_{1212}^{\text{\tiny{S}}}[x;x']=\frac{1}{2}\,\sqrt{{\gamma_{11}(x)}{\gamma_{22}(x)}}\sqrt{{\gamma_{11}(x')}{\gamma_{22}(x')}} \, {\cos \chi(v,v')}\, ,
        \\
        & \hspace{-0.9 cm}\bohm_{2212}^{\text{\tiny{S}}}[x;x']=\frac{1}{2}\,\sqrt{\frac{\gamma_{22}(x)}{\gamma_{11}(x)}}\sqrt{{\gamma_{11}(x')}{\gamma_{22}(x')}}\ (-\sin\chi(v,v')+\gamma_{12}(x)\cos\chi(v,v'))\, ,\\
        & \hspace{-0.9 cm}\bohm_{2222}^{\text{\tiny{S}}}[x;x']=\frac{1}{2}\,\sqrt{\frac{\gamma_{22}(x)}{\gamma_{11}(x)}}\sqrt{\frac{\gamma_{22}(x')}{\gamma_{11}(x')}}\left[(\gamma_{12}(x)-\gamma_{12}(x'))\sin\chi(v,v')+(1+\gamma_{12}(x)\gamma_{12}(x'))\cos\chi(v,v')\right]\, ,
    \end{align}
\end{subequations}
where 
\begin{equation}
    \chi(v,v')=-\frac12\int_{v'}^v \d{}\tilde{v}\ \gamma_{12}(\tilde{v})\left(\frac{\partial_{\tilde{v}}\gamma_{11}(\tilde{v})}{\gamma_{11}(\tilde{v})}-\frac{\partial_{\tilde{v}}\gamma_{22}(\tilde{v})}{\gamma_{22}(\tilde{v})}\right)\, .
\end{equation}

\addcontentsline{toc}{section}{References}
\bibliographystyle{fullsort.bst}
\bibliography{reference}

\providecommand{\href}[2]{#2}\begingroup\raggedright\endgroup

\end{document}